\documentclass[journal,twoside,web]{ieeecolor}
\usepackage{generic}
\usepackage{cite}
\usepackage{amsmath,amssymb,amsfonts}
\usepackage{graphicx}
\usepackage{textcomp}

\usepackage{bm}
\usepackage{mathrsfs}
\usepackage{color}
\usepackage{xcolor}
\usepackage{url}
\usepackage{longtable}
\usepackage{multirow}
\usepackage{algorithm}
\usepackage{algorithmicx}
\usepackage{algpseudocode}
\usepackage{multirow}
\algrenewcommand{\algorithmiccomment}[1]{\hskip3em\% #1}

\usepackage{dblfloatfix}

\usepackage[hang,small,bf]{caption}
\usepackage[subrefformat=parens]{subcaption}
\newtheorem{assumption}{Assumption}
\newtheorem{lemma}{Lemma}
\newtheorem{theorem}{Theorem}

\newtheorem{proposition}{Proposition}
\newtheorem{problem}{Problem}
\newtheorem{remark}{Remark}

\newcommand{\A}{\mathcal{A}}

\newcommand{\I}{\mathcal{I}}
\newcommand{\J}{\mathcal{J}}
\newcommand{\K}{\mathcal{K}}

\newcommand{\cO}{\mathcal{O}}

\newcommand{\cP}{\mathcal{P}}

\newcommand{\R}{\mathbb{R}}

\newcommand{\Z}{\mathbb{Z}}

\newcommand{\im}{\mathrm{im\hspace{0.2ex}}}
\newcommand{\rank}{\mathrm{rank\hspace{0.2ex}}}
\newcommand{\supp}[1]{{\rm supp}\left({#1}\right)}

\def\BibTeX{{\rm B\kern-.05em{\sc i\kern-.025em b}\kern-.08em
		T\kern-.1667em\lower.7ex\hbox{E}\kern-.125emX}}
\begin{document}
	\title{Detection and Identification of Sensor Attacks Using {Partially Attack-Free} Data}
	
	\author{Takumi Shinohara, \IEEEmembership{Member, IEEE}, Karl Henrik Johansson, \IEEEmembership{Fellow, IEEE}, \\ and Henrik Sandberg, \IEEEmembership{Fellow, IEEE}
		\thanks{This work was supported in part by the Knut and Alice Wallenberg Foundation Wallenberg Scholar Grant, Swedish Research Council (Project 2023-04770), and VINNOVA project ``Control-computing-communication co-design for autonomous industry (3C4AI)'' (No. 2025-01119).}
		\thanks{The authors are with the Division of Decision and Control Systems, KTH Royal Institute of Technology, and also with Digital Futures, 100 44 Stockholm, Sweden. (e-mail: tashin@kth.se, kallej@kth.se, hsan@kth.se).}}
	
	\maketitle
	
	\begin{abstract}
		In this paper, we investigate data-driven attack detection and identification in a model-free setting.
		{We consider a practically motivated scenario in which the available dataset may be compromised by malicious sensor attacks, but contains an unknown, contiguous, partially attack-free interval.
			The control input is assumed to include a small stochastic watermarking signal.
			Under these assumptions, we establish sufficient conditions for attack detection and identification from partially attack-free data.
			We also develop data-driven detection and identification procedures and characterize their computational complexity.
			Notably, the proposed framework does not impose a limit on the number of compromised sensors; thus, it can detect and identify attacks even when all sensor outputs are compromised outside the attack-free interval, provided that the attack-free interval is sufficiently long.}
		Finally, we demonstrate the effectiveness of the proposed framework via numerical simulations.
	\end{abstract}
	
	
	\section{Introduction}
	\label{sec:introduction}
	\IEEEPARstart{U}{biquitous} network connectivity and data exchange have made control systems more vulnerable to malicious cyberattacks.
	To address this issue, numerous efforts toward attack detection and identification have been made to enhance system security and resilience.
	Most existing studies (e.g., \cite{2013TACBullo,2015Johansson,2019TACShinohara}) rely on prior knowledge of the system's mathematical model.
	However, deriving precise mathematical models can be hard or even infeasible.
	Consequently, data-driven approaches are important to ensure secure and resilient operations, complementing traditional model-based methods.
	
	The paper \cite{2023TACShi} presents data-driven methods for attack detection and identification based on subspace identification and $ \ell_2/\ell_1 $ optimization frameworks. 
	In \cite{2025ECCTeixeira}, the authors develop a data-driven attack detection algorithm in networked control systems and provide corresponding feasibility conditions.
	The paper \cite{2025TCS-ILi} proposes an iterative reweighted $ \ell_2/\ell_1 $ minimization approach to enhance the performance of data-driven attack detection and identification.
	The paper \cite{2021TCYBLi} investigates a data-driven design scheme for undetectable false data-injection attacks and proposes a detection scheme that leverages coding theory.
	The secure data reconstruction problem under data manipulation, using the behavioral approach, is addressed in \cite{2026TACLygeros}.
	These studies assume that historical datasets are free from malicious sensor attacks and thus build their frameworks on clean (i.e., attack-free) datasets.
	In practice, however, the data may already be corrupted by malicious false-data injection attacks.
	
	{The paper \cite{2021L-CSSPasquialetti} considers data-driven attack detection using potentially compromised data, without requiring model identification.
		A key concept in \cite{2021L-CSSPasquialetti} is a \textit{safe time} after which detection capability against a specific class of attacks is guaranteed; attacks occurring before the safe time are undetectable, and the analysis therefore focuses on attacks starting after the safe time.
		In other words, an initial attack-free period (from time zero to the safe time) is required to learn the system model.
		Moreover, \cite{2021L-CSSPasquialetti} focuses on attack detection only.
		While detection alone can raise an alarm, identifying which sensors are compromised is crucial for isolation and for recovering a usable dataset for subsequent data-driven control.}
	
	{
		In this paper, we study both attack detection and identification using known input data and potentially compromised output measurements, without assuming a system model.
		Instead of postulating an initial safe time, we consider a practically motivated setting where the available dataset is compromised but contains an unknown contiguous attack-free interval.
		This reflects operational scenarios in which an adversary's ability to manipulate sensor data is intermittent (e.g., due to re-authentication, software updates, or maintenance).
		We further assume that the control input incorporates a small i.i.d. Gaussian watermarking signal to improve attack detection and identification performance.
		Such a random signal is a common practice in system identification and data-driven control, and can be made small enough to have a negligible impact on nominal operation.
		Under these assumptions, we derive data-driven sufficient conditions for detecting and identifying attacks and establish corresponding algorithms.
		Our primary contributions can be summarized as follows:
		\begin{enumerate}
			\item We show that data-driven attack detection is feasible using partially attack-free data, provided that the attack-free interval is sufficiently long. Based on the attack-detection condition, we propose a heuristic attack-detection algorithm.
			\item We develop an attack-identification condition based on singular value decomposition (SVD), and show that the identification is also feasible when the attack-free interval is sufficiently long. Given the condition, we propose an identification algorithm to uniquely recover the compromised sensor set.
			\item We quantify the computational complexity of the proposed attack detection and identification algorithms.
		\end{enumerate}
		It is worth noting that our attack detection and identification algorithms do not impose a limit on the number of compromised sensors; thus, they remain feasible even when all sensor outputs are compromised, except during the attack-free interval, provided that this interval is sufficiently long.
	}
	
	{
		The rest of this paper is organized as follows.
		Section~\ref{section:problem} introduces the system model, input design, data representation of the system, and assumption of partially attack-free data.
		In Section~\ref{section:detection}, we address the detection problem, and we derive the data-driven detection condition and propose a corresponding heuristic algorithm based on partially attack-free data.
		Section~\ref{section:identification} is devoted to deriving the attack identification condition and providing a procedure to identify the compromised sensor set from partially attack-free data.
		In Section~\ref{section:simulation}, we present simulation results to show the validity of our framework.
		Section~\ref{section:conclusion} finally concludes this paper.
	}

	\subsubsection*{Notations}
	The symbols $ \R $, $ \R^n$, and $ \mathbb{Z}^+ $ denote the set of real numbers, $ n $-dimensional Euclidean space, and positive integers, respectively.
	The notation $ |\I| $ denotes the cardinality of a set $ \I $.
	For a vector $ x $, its support is defined as $ \supp{x} $.
	Given a linear map $ A $, we use $ \ker A $ and $ \im A $ to denote the kernel and image of $ A $, respectively.
	The identity matrix of dimension $ n \times n $ is denoted as $ I_n $.
	Given two sets $ \mathcal{U} $ and $ \mathcal{W} $, the Minkowski sum is defined by $ \mathcal{U} + \mathcal{W} \triangleq \{u+w:u\in\mathcal{U}, w \in \mathcal{W}\} $.
	For $ a, b \in \mathbb{Z}^+ $, we define the integer interval $ [a,b] \triangleq \{c \in \Z^+:a \leq c \leq b \} $, where $ [a,b] = \emptyset $ if $ a > b $.
	Given a sequence $ \{v(0),\ldots,v(N-1)\} $ and the interval $ [i,j] $ with $ i \geq 0 $, $ j \leq N-1 $, and $ i < j $, we define
	\begin{align}
		\label{eq:v^[i,j]}
		v^{[i,j]} \triangleq \left[~v(i)^\top~v(i+1)^\top~\cdots~v(j-1)^\top~v(j)^\top~\right]^\top.
	\end{align}
	We sometimes use the simple notation $ v^{[j]} $ instead of $ v^{[0,j]} $.
	{
		Let $ q $ be a positive integer such that $ q \leq j - i + 1 $ and define the Hankel matrix of depth $ q $, associated with $ v^{[i,j]} $, as
		\begin{align*}
			\mathscr{H}_q\!\left(\!v^{[i,j]}\!\right) \!\triangleq \!\left[\!\!
			\begin{array}{cccc}
				v(i) & v(i\!+\!1) & \!\cdots \!& v(j\!-\!q\!+\!1) \\
				v(i\!+\!1) & v(i\!+\!2) &\! \cdots \!& v(j\!-\!q\!+\!2) \\
				\vdots & \vdots & \!\ddots \!& \vdots \\
				v(i\!+\!q\!-\!1) & v(i\!+\!q) & \!\cdots\! & v(j)
			\end{array}\!\!\right]\!.
		\end{align*}
		Note that the subscript $ q $ refers to the number of block rows of the Hankel matrix.
		Then, $ v^{[i,j]} $ is said to be \textit{persistently exciting of order $ q $} if the Hankel matrix $ \mathscr{H}_q(v^{[i,j]}) $ has full row rank \cite{2005SCLWillems}.
		Under the persistently exciting input condition, Willems' fundamental lemma implies that the Hankel matrix constructed from (attack-free) data spans the system's behavior and thus determines its left-kernel representation (for details, see, e.g., \cite{2005SCLWillems,Waarde_Data-Driven}).
		Our proposed framework will exploit this rank/left-kernel structure to detect and identify attacks in data that may be compromised but contain an unknown contiguous attack-free interval.}
	
	{
		For random variables (and vectors/matrices) $ x $ and $ y $, we use the notation $ x \overset{\mathrm{a.s.}}{=} y $ for almost-sure equality, i.e., $ \mathbb{P}(x = y) = 1 $.
		Similarly, we use the notation $ x \overset{\mathrm{a.s.}}{\neq} y $ if $ \mathbb{P}(x \neq y) = 1 $.}


\section{Problem Formulation}
\label{section:problem}
In this section, we introduce the system model and {the input design}.
We then describe the data representation used in this paper and {the partially attack-free data assumption}.

\subsection{System Model and Input Design}
We consider the following discrete-time linear time-invariant system subject to malicious sensor attacks:
\begin{align}
	\label{eq:sytem_model}
	\left\lbrace
	\begin{array}{rl}
		x(k+1) \!\!\!&= Ax(k) + Bu(k), \\
		y(k) \!\!\! &= Cx(k) + a(k),
	\end{array} \right.
\end{align}
where $ x(k) \in \R^n $ is the unknown system state, $ u(k) \in \R^m $ is the control input, and $ y(k) \in \R^p $ is the possibly compromised system output.
The vector $ a(k) \in \R^p $ stands for sensor attacks designed by a malicious attacker.
For notational convenience, define $\mathcal{P} \triangleq \{1,\ldots,p\}$ as the index set of the sensors.
For the system, we have the following assumption, which is standard in data-driven problems.
\begin{assumption}
	\label{assumption:system}
	The system is controllable and observable. The system parameters $ A $, $ B $, and $ C $ are unknown, but the input and output (I/O) data from time $ 0 $ to $ N-1 $ of (\ref{eq:sytem_model}), namely $ u^{[N-1]} $ and $ y^{[N-1]} $ are known, where $ N $ is a sufficiently large integer.
\end{assumption}

{In this paper, we assume that the control input incorporates an additive watermarking signal.}
\begin{assumption}
	\label{assumption:input}
	{The control input is designed as
		\begin{align}
			\label{eq:control}
			u(k) = u_{\rm nom}(k) + w(k),
		\end{align}
		where $ \{u_{\rm nom}(k)\} $ is the nominal input sequence and $ \{w(k)\} $ is an i.i.d. Gaussian watermark sequence with zero mean and covariance $ \varphi^2 I_m $.
		The watermark $ w(k) $ is independent of the past and current nominal input sequence $ \{u_{\rm nom}(t)\}_{t=0}^{k} $.
	}
\end{assumption}

{
	This assumption is introduced for two complementary reasons: (i) input excitation and (ii) security enhancement.
	From a system identification and data-driven control perspective, many methods require input excitation, which is expressed as the full-row-rank condition of the input Hankel matrix (see, e.g., \cite{2005SCLWillems,System Identification,2020TACTrentelman,Waarde_Data-Driven}).
	A random input is a standard way to ensure that the persistent excitation of arbitrary order is satisfied almost surely \cite{2023LCSSAlsalti,2026TACOzay}.
	Indeed, some data-driven control applications adopt such random signals to construct a stable data-driven controller (see, e.g., \cite{2021Dorfler,2023ARCLazar}).
	From a security perspective, injecting watermarking signals into the control input is a well-established defense strategy against malicious sensor attacks (see, e.g., \cite{2014TCSTSinopoli,2020TACJohansson}).
	There is a trade-off between control performance and attack detectability/identifiability depending on the covariance $ \varphi^2 $.
	In practice, this covariance can be chosen to be small enough to preserve nominal closed-loop performance, while still providing the excitation required for the rank-based analysis.
}

For the attacker and the attack signal, we have the following assumption.
\begin{assumption}
	\label{assumption:attack}
	The adversary possesses knowledge of the system's state, control input, sensor measurements, and system model.
	The attack signal $ a(k) $ is arbitrarily designed by the adversary, {i.e., $ a(k) $ can be designed based on the past and current state sequence $ \{x(t)\}_{t=0}^{k} $, the input sequence $ \{u(t)\}_{t=0}^{k-1} $, and the output sequence $ \{y(t)\}_{t=0}^{k-1} $}.
	Define the number of compromised sensors as $ \ell $, and the subset of compromised sensors is fixed over time, denoted by $ {\mathcal{A}^*} \subseteq \cP $ with $ |{\mathcal{A}^*}| = \ell$, i.e., $ \supp{a(k)} = {\A^*} $ if $ a(k) \neq 0 $.
	The number of compromised sensors $ \ell $ and the compromised sensor set $ {\mathcal{A}^*}  $ are unknown to the system operator.
\end{assumption}

{
	Unlike generic disturbances, attacks are strategically designed by adversaries.
	This allows them to select injected signals that mimic legitimate trajectories and evade detection.
	Such adversarial characteristics make detection and identification significantly more challenging using conventional anomaly detectors.}

	\subsection{Data Representation}
	For a given positive integer $ q \leq N $, we obtain the following stacked observation model from (\ref{eq:sytem_model}):
	\begin{align}
		\label{eq:collected_measure}
		y^{[k,k+q-1]} =\mathcal{O}^q x(k) \!+\! \mathcal{T}^qu^{[k,k+q-1]} \!+\! a^{[k,k+q-1]},
	\end{align}
	{where $ y^{[k,k+q-1]} \in \R^{pq} $ is the stacked vector containing the values of $ y $ from time $ k $ to $ k+q-1 $, following the notation (\ref{eq:v^[i,j]}).
		The vectors $ u^{[k,k+q-1]} \in \R^{mq} $ and $ a^{[k,k+q-1]} \in \R^{pq} $ are similarly defined from $ u $ and $ a $, respectively.}
	The matrix $ \cO^q $ is the extended observability matrix
	\begin{align*}
		\cO^q \triangleq \left[
		\begin{array}{cccc}
			C^\top & (CA)^\top & \cdots & (CA^{q-1})^\top
		\end{array}\right]^\top \in \R^{pq \times n},
	\end{align*}
	and $ \mathcal{T}^q $ is the input-to-output Toeplitz matrix
	\begin{align*}
		\mathcal{T}^q \triangleq \left[
		\begin{array}{cccc}
			0 & 0 & \cdots & 0 \\
			CB & 0 & \cdots & 0 \\
			\vdots & \ddots & \ddots & \vdots \\
			CA^{q-2}B & \cdots & CB & 0
		\end{array}\right]\in \R^{pq \times mq}.
	\end{align*}
	The Hankel matrix associated with \textit{all} output data $ y^{[N-1]} $ is denoted as
	\begin{align*}
		&Y^q \triangleq {\mathscr{H}_q\left(y^{[N-1]}\right) } \nonumber \\
		&{=\!\left[\!\!
			\begin{array}{cccc}
				y(0) & y(1) & \!\cdots \!& y(N\!-\!q\!+\!1) \\
				y(1) & y(2) &\! \cdots \!& y(N\!-\!q) \\
				\vdots & \vdots & \!\ddots \!& \vdots \\
				y(q\!-\!1) & y(q) & \!\cdots\! & y(N\!-\!1)
			\end{array}\!\!\right]}  \in  \R^{pq  \times  (N-q+1)}.
	\end{align*}
	Then, we have the following input and output model:
	\begin{align}
		\label{eq:IO_data_model}
		Y^q =  \mathcal{O}^q X + \mathcal{T}^q U^q + \Lambda^q,
	\end{align}
	where
	\begin{align*}
		X &\triangleq  \left[
		\begin{array}{ccc}
			x(0) & \cdots & x(N-q)
		\end{array}\right] \in \R^{n \times (N-q+1)}, \\
		U^q &\triangleq {\mathscr{H}_q\left(u^{[N-1]}\right) \in \R^{mq \times (N-q+1)}} , \\
		\Lambda^q &\triangleq {\mathscr{H}_q\left(a^{[N-1]}\right) \in \R^{pq \times (N-q+1)}}.
	\end{align*}
	For future analysis, we define {the stacked I/O vector from time $ k $ to $ k+q-1 $ as}
	\begin{align}
		\label{eq:z_vector}
		z^{[k,k+q-1]} & \triangleq \left[
		\begin{array}{c}
			u^{[k,k+q-1]} \\ y^{[k,k+q-1]}
		\end{array}\right]\in \R^{mq+pq}.
	\end{align}

\begin{figure}[t]
		\begin{center}
			\includegraphics[width=\linewidth]{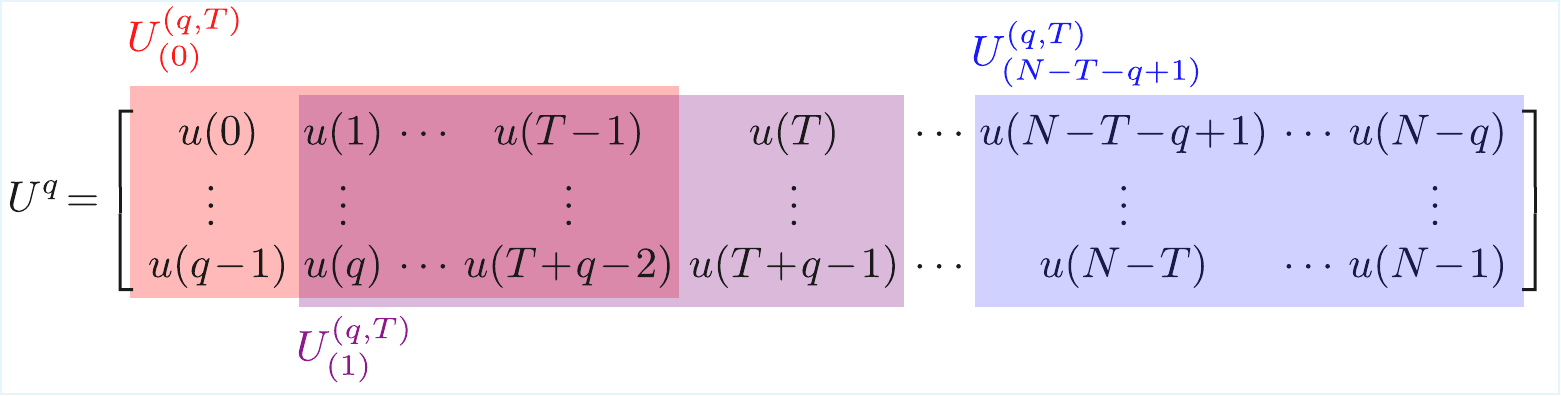}
			\vspace{-7.5mm}
			\caption{Relationship between $ U^q $ and $ U^{(q,T)}_{(k)} $. The same applies to $ Y $, $ \Lambda $, and $ Z $.}
			\label{fig:Relationship_U}
		\end{center}
		\vspace{-4mm}
	\end{figure}

	Define the Hankel matrix associated with the output data in the \textit{time interval $ [k, k+T+q-2] $ with length $ T\in[1,N-q+1] $} as
	\begin{align*}
		& Y^{(q,T)}_{(k)} \triangleq {\mathscr{H}_q\left(y^{[k,k+T+q-2]}\right)} \nonumber \\
		&{=\!\left[\!\!
			\begin{array}{cccc}
				y(k) & y(k\!+\!1) & \!\cdots \!& y(k\!+\!T\!-\!1) \\
				y(k\!+\!1) & y(k\!+\!2) &\! \cdots \!& y(k\!+\!T) \\
				\vdots & \vdots & \!\ddots \!& \vdots \\
				y(k\!+\!q\!-\!1) & y(k\!+\!q) & \!\cdots\! & y(k\!+\!T\!+\!q\!-\!2)
			\end{array}\!\!\right]} \!\! \in \! \R^{pq \times T}.
	\end{align*}
	Similarly, denote
	\begin{align*}
		X^{(T)}_{(k)} &\triangleq \left[
		\begin{array}{ccc}
			x(k) & \cdots & x(k+T-1)
		\end{array}\right] \in \R^{n \times T} , \\
		U^{(q,T)}_{(k)} &\triangleq {\mathscr{H}_q\left(u^{[k,k+T+q-2]}\right) \in \R^{mq \times T}} , \\
		\Lambda^{(q,T)}_{(k)} &\triangleq {\mathscr{H}_q\left(a^{[k,k+T+q-2]}\right) \in \R^{pq \times T}}.
	\end{align*}
	Then, we have the following input and output model for time $ k $ with length $ T $:
	\begin{align}
		\label{eq:Y_q,T}
		Y^{(q,T)}_{(k)} &= \mathcal{O}^q X^{(T)}_{(k)} + \mathcal{T}^q U^{(q,T)}_{(k)} + \Lambda^{(q,T)}_{(k)}.
	\end{align}
	The I/O data matrix for time $ k $ with length $ T $ is defined as
	\begin{align}
		\label{eq:Z_matrix_k}
		Z^{(q,T)}_{(k)}&\triangleq\left[
		\begin{array}{c}
			U^{(q,T)}_{(k)}\\ Y^{(q,T)}_{(k)}
		\end{array}\right]\in \R^{(mq+pq) \times T}.
	\end{align}
	The relationship between the matrices $ \bullet^q $ and $ \bullet^{(q,T)}_{(k)} $ can be illustrated in Fig.~\ref{fig:Relationship_U}, where $ \bullet $ is $ Y, U, \Lambda $, or $ Z $.
	
	\subsection{Partially Attack-Free Data}
	In this paper, we focus on the data-driven attack detection and identification analysis.
	If the attacker can inject malicious signals into the data over the entire time span (i.e., from $ 0 $ to $ N-1 $), they can design undetectable attacks against the system, as shown in \cite{2021L-CSSPasquialetti,2021TCYBLi,2023TACShi}.
	Hence, we make the following assumption regarding the presence of an \textit{attack-free interval} for the analysis.
	\begin{assumption}
		\label{assumption:clear_data}
		In the given I/O data, there exists an \textit{attack-free interval} $ \K_0 \triangleq [k_0, k_0 + \tau - 1] \subseteq [0, N-1] $ of length $ \tau $ during which $ a(k) = 0 $ for all $ k \in \K_0 $.
		This interval $ \K_0 $ is maximal within $ [0,N-1] $, i.e., if $ k_0 > 0 \Rightarrow a(k_0 -1) \neq 0$ and $k_0+\tau -1 < N-1 \Rightarrow  a(k_0 +\tau) \neq 0 $.
	\end{assumption}
	
	\begin{figure}[t]
		\begin{center}
			\includegraphics[width=\linewidth]{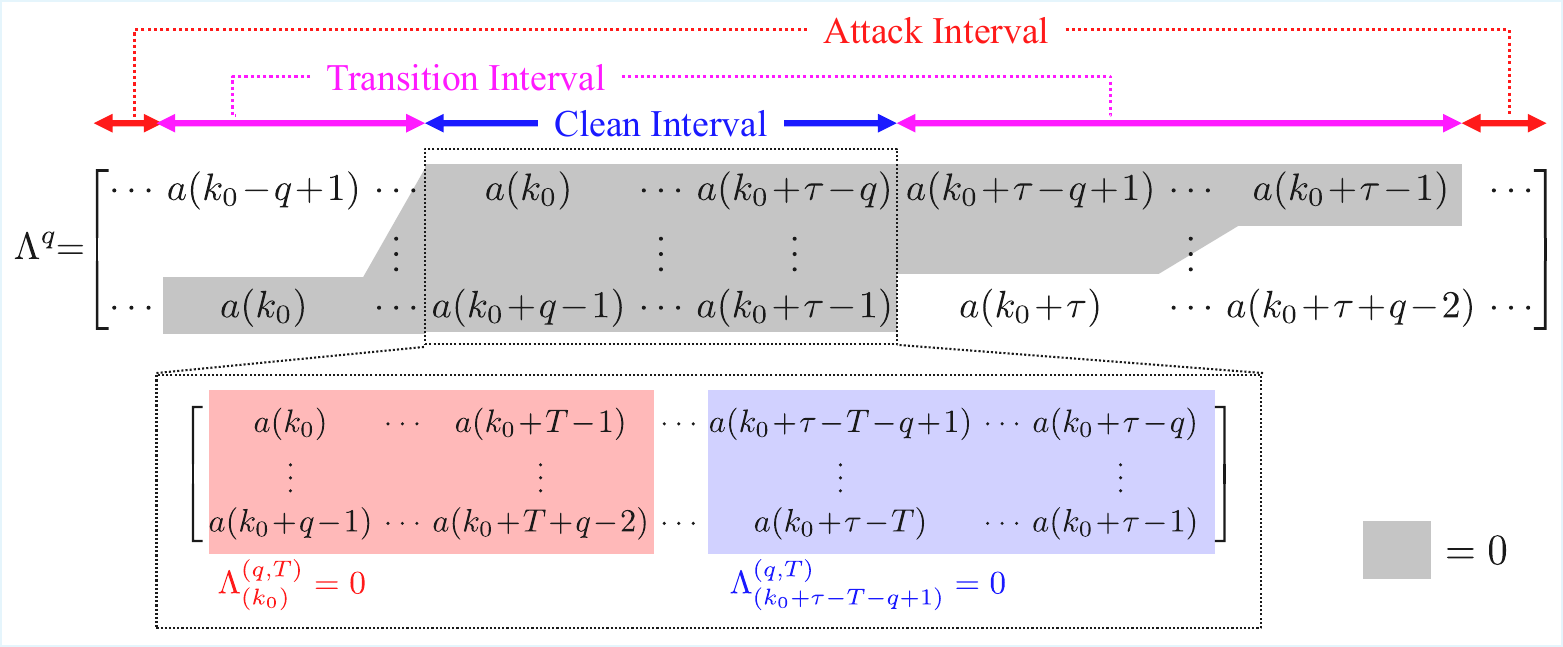}
			\vspace{-7mm}
			\caption{Graphical explanation of $ \Lambda^q $ and $ \Lambda^{(q,T)}_{(k)} $ considering the attack-free interval $ \K_0 = [k_0,k_0+\tau-1]$, where the gray-shaded elements are all zero, and the clean, transition, and attack intervals are illustrated in the vector sense.}
			\vspace{-4mm}
			\label{fig:Relationship_Lambda}
		\end{center}
	\end{figure}
	
	This assumption implies that the output $ y(k) $ for $ k \in \K_0 $ is immune to sensor attacks, i.e., the data are partially attack-free.
	When such an interval $ \K_0 $ exists in the dataset, the graphical explanation of $ \Lambda^q $ and $ \Lambda^{(q,T)}_{(k)} $ can be depicted in Fig.~\ref{fig:Relationship_Lambda}, where, for a vector $ \blacktriangle^{[k, k+q-1]} $, we classify it as being in the
	\begin{align*}
		\left\lbrace
		\begin{array}{ll}
			\mathrm{clean~interval}, & \mathrm{if}~k \in [k_0,k_0+\tau - q],\\
			\mathrm{transition~interval}, & \mathrm{if}~k \in [k_0-q+1,k_0-1] \\
			&\hspace{2mm}~\mathrm{or}~ [k_0+\tau-q+1,k_0+\tau-1],\\
			\mathrm{attack~interval}, & \mathrm{otherwise},\\
		\end{array}\right. \!\!
	\end{align*}
	where $ \blacktriangle $ is $ y, u, a $, or $ z $.
	Also, for a matrix $ \bullet^{(q,T)}_{(k)} $, we classify it as being in the
	\begin{align*}
		\left\lbrace
		\begin{array}{ll}
			\mathrm{clean~interval}, & \mathrm{if}~k \in [k_0,k_0+\tau - T - q + 1],\\
			\mathrm{transition~interval}, & \mathrm{if}~k \in [k_0-T-q+2,k_0-1] \\
			&\hspace{2mm}~\mathrm{or}~ [k_0\!+\!\tau \!-\! T \!-\! q \!+\! 2,k_0\!+\!\tau\!-\!1],\\
			\mathrm{attack~interval}, & \mathrm{otherwise},\\
		\end{array}\right.\!\!\!
	\end{align*}
	where $ \bullet $ is $ Y,U, \Lambda $, or $ Z $.
	
	Note that we now assume the existence of $ \K_0 $, but its location and length are unknown.
	In other words, the system operator knows that there is an attack-free interval in the output data, but does not know which data are clean and for how long.
	Thus, it should also be noted that for some $ \blacktriangle^{[k,k+q-1]} $ and $ \bullet^{(q,T)}_{(k)} $, we cannot determine whether these are in the clean, transition, or attack interval.
	
	\begin{remark}
		\label{remark:clean_data}
		In general, it is difficult to know the existence of a clean interval.
		However, the likelihood of having such an interval can be increased by collecting a substantial amount of data, taking into account the adversary's capabilities and technical constraints.
	\end{remark}

	The goal of the system operator is to detect and identify sensor attacks using compromised I/O data that contain a partially attack-free interval.
	This is formally defined as follows.
	\begin{problem}[Data-Driven Attack Detection]
		\label{problem:attack_detector}
		Given the I/O data $ u^{[N-1]} $ and $ y^{[N-1]} $ {with a partially attack-free interval}, determine if the sensor attack $ a(k) \neq 0 $ exists for some $ k \in [0,N-1] $.
	\end{problem}
	\begin{problem}[Data-Driven Attack Identification]
		\label{problem:sensor_identifier}
		Given the I/O data $ u^{[N-1]} $ and $ y^{[N-1]} $ {with a partially attack-free interval}, identify the unique set of compromised sensors $ {\mathcal{A}^*} \subseteq \cP$.
	\end{problem}

\section{Data-Driven Attack Detection with Partially Attack-Free Data}
\label{section:detection}
{In this section, we consider Problem~\ref{problem:attack_detector}.
	We first introduce some important rank conditions to be used, followed by the concrete detection condition and the corresponding algorithm.}

\subsection{Preliminaries}
{Under the input design of (\ref{eq:control}), the partial input data matrix $ U^{(q,T)}_{(k)} $ can have full row rank with probability one, as provided in the following lemma.
	\begin{lemma}
		\label{lemma:U_k}
		Suppose that Assumption \ref{assumption:input} holds, i.e., the control input $ u(k) $ is designed as (\ref{eq:control}).
		Fix any positive integers $ q \in[1,N-T+1]$ and $ T \in [mq,N-q+1] $.
		Then
		\begin{align}
			\rank U^{(q,T)}_{(k)}  \overset{\mathrm{a.s.}}{=} mq,~\forall k \in [0,N-T-q+1].
		\end{align}
	\end{lemma}
	\vspace{1.5mm}
	\begin{proof}
		See Appendix~\ref{appendix:proof_lemma_U}.
\end{proof}}

{
	This lemma implies that, under the input design of (\ref{eq:control}), for any integers $ q \in[1,N-T+1]$ and $ T \in [mq,N-q+1] $, the Hankel matrix $ U^{(q,T)}_{(k)} $ has full row rank for all $ k $ with probability one.
	Equivalently, every input window $ u^{[k,k+T+q-2]} $ is persistently exciting of order $ q $ for all $ k $ almost surely.}

{
	We also have the following lemma that states the rank condition of $ X^{(T)}_{(k)} $.
	\begin{lemma}
		\label{lemma:X_rank}
		Suppose that Assumptions \ref{assumption:system} and \ref{assumption:input} hold, i.e., the system is controllable and the control input $ u(k) $ is designed as (\ref{eq:control}).
		If $ T \in [n+1,N-1]$, then
		\begin{align}
			\rank X^{(T)}_{(k)}  \overset{\mathrm{a.s.}}{=} n,~\forall k \in [0,N-T].
		\end{align}
	\end{lemma}
	\vspace{1.5mm}
	\begin{proof}
		See Appendix~\ref{appendix:proof_lemma_X}.
\end{proof}}

{We observe that if the input $ u(k) $ is designed as (\ref{eq:control}), $ X^{(T)}_{(k)} $ has full row rank for all $ k $ with $ T \in [n+1,N-1] $ almost surely.
	This property is valuable for detecting and identifying attacks when relying solely on partially attack-free data.
}

{
	Additionally, we introduce the following lemma on the rank of the stacked matrix of $ U^{(q,T)}_{(k)} $ and $ X^{(T)}_{(k)} $.
	\begin{lemma}
		\label{lemma:U-X_rank}
		Suppose the same assumptions as in Lemma~\ref{lemma:X_rank}.
		Fix any positive integers $ q \in [1,N-T+1] $ and $ T \in [mq + n + 1, N-q+1]$.
		Then,
		\begin{align}
			\label{eq:lemma_U-X_rank}
			\rank \!\left[\!
			\begin{array}{c}
				U^{(q,T)}_{(k)} \\ X^{(T)}_{(k)}
			\end{array}\!\right]  \overset{\mathrm{a.s.}}{=} mq + n,~\!\forall k \in [0,N-T-q+1].
		\end{align}
	\end{lemma}
	\vspace{1.5mm}
	\begin{proof}
		This lemma can be proved using the same polynomial-measure argument as Lemmas \ref{lemma:U_k}--\ref{lemma:X_rank}, and thus we omit the details.
\end{proof}}

\subsection{Attack Detection Condition}
{Using the rank properties from the previous subsection, we derive the attack-detection condition using partially attack-free data.
	First, we introduce the following proposition on the rank of $ Z^{(q,T)}_{(k)} $ in an attack-free scenario.
	\begin{proposition}
		\label{proposition:detection_attack-free}
		Suppose that Assumptions \ref{assumption:system} and \ref{assumption:input} hold, i.e., the system is controllable and observable, $ A,B,C $ are unknown, the I/O data are given, and $ u(k) $ is designed by (\ref{eq:control}).
		Also, assume that $ a(k) \equiv 0 $ for all $ k \in [0,N-1] $.
		Fix any integers $ q \in [1, N-T+1] $ and $ T \in [mq + n + 1, N-q+1]  $.
		Then
		\begin{align}
			\label{eq:lemma_Z_rank}
			\rank Z^{(q,T)}_{(k)} \overset{\mathrm{a.s.}}{=} mq \!+\!\rank \mathcal{O}^q,~\forall k \in [0,N\!-\!T\!-\!q\!+\!1].
		\end{align}
		Specifically, if $ q \in [n, N-T+1] $ and $ T \in [mq + n + 1, N-q+1]  $, then
		\begin{align}
			\label{eq:lemma_Z_rank_n}
			\rank Z^{(q,T)}_{(k)} \overset{\mathrm{a.s.}}{=} mq \!+\! n,~\forall k \!\in\! [0,N\!-\!T\!-\!q\!+\!1].
		\end{align}
\end{proposition}}
\vspace{1.5mm}
\begin{proof}
	For (\ref{eq:Z_matrix_k}), since $ a(k) \equiv 0 $ for all $ k \in [0,N-1] $, we can ignore the attack matrix $ \Lambda^{(q,T)}_{(k)} $ and consider the rank of
	\begin{align*}
		Z^{(q,T)}_{(k)} = \underbrace{\left[
			\begin{array}{c}
				U^{(q,T)}_{(k)} \\ \mathcal{T}^q U^{(q,T)}_{(k)}
			\end{array}\right]}_{L^1_{(k)}}  + \underbrace{\left[
			\begin{array}{c}
				0 \\ \mathcal{O}^q  X^{(T)}_{(k)}
			\end{array}\right]}_{L^2_{(k)}}.
	\end{align*}
	For two sets $ \mathcal{U} $ and $ \mathcal{W} $, the following formula is well known:
	\begin{align}
		\label{eq:Grassmann_formula}
		\dim(\mathcal{U} + \mathcal{W}) + \dim(\mathcal{U} \cap \mathcal{W}) = \dim \mathcal{U} + \dim \mathcal{W}.
	\end{align}
	Applying this formula to two sets $ \im  L^1_{(k)}$ and $ \im  L^2_{(k)} $, we obtain
	\begin{align}
		\label{eq:L^1_and_L^2_rank}
		& \dim(\im  L^1_{(k)} + \im  L^2_{(k)}) + \dim(\im  L^1_{(k)}\cap \im  L^2_{(k)})  \nonumber\\
		&= \dim \im  L^1_{(k)}+ \dim \im L^2_{(k)} = \rank L^1_{(k)} + \rank L^2_{(k)}
	\end{align}
	for all $ k \in [0,N-T-q+1]$.
	
	We first show $ \im Z^{(q,T)}_{(k)} = \im  L^1_{(k)} + \im  L^2_{(k)} $.
	{Since $ q \in [1, N-T+1] $ and $ T \in [mq + n + 1, N-q+1]  $, from Lemma~\ref{lemma:U-X_rank}, (\ref{eq:lemma_U-X_rank}) holds.
		Thus, for any $ u \in \R^{mq} $ and $ x \in \R^n $, there exists $ \alpha \in \R^T $ such that  $ U^{(q,T)}_{(k)}  \alpha = u $ and $  X^{(T)}_{(k)} \alpha = x $ for all $ k \in [0, N-T-q+1] $ almost surely.
		Hence, we obtain
		\begin{align*}
			Z^{(q,T)}_{(k)} \alpha \!= \!\left[\!\!
			\begin{array}{c}
				U^{(q,T)}_{(k)} \\ \mathcal{T}^q U^{(q,T)}_{(k)}
			\end{array}\!\!\right]\!\alpha\! +\!\left[\!\!
			\begin{array}{c}
				0 \\ \mathcal{O}^q  X^{(T)}_{(k)}
			\end{array}\!\!\right]\!\alpha \!\overset{\rm a.s.}{=}\!  \left[\!\!
			\begin{array}{c}
				u \\ \mathcal{T}^q u
			\end{array}\!\!\right] \!+\!\left[\!\!
			\begin{array}{c}
				0 \\ \cO^q x
			\end{array}\!\!\right]\!\!
		\end{align*}
		for some $ \alpha \in \R^T $, $ u \in \R^{mq} $, and $ x \in \R^{n} $.}
	By construction, we have
	\begin{align*}
		\im L^1_{(k)} \!=\! \left\lbrace \left[\!
		\begin{smallmatrix}
			I_{mq} \\ \mathcal{T}^q
		\end{smallmatrix}\!\right]\!u \!:\!u \in \R^{mq} \right\rbrace\!,~\im L^2_{(k)} \!=\! \left\lbrace \left[\!
		\begin{smallmatrix}
			0 \\ \mathcal{O}^q
		\end{smallmatrix}\!\right]x\!:\!x\in \R^n\right\rbrace\!,
	\end{align*}
	which implies the relation $ \im Z^{(q,T)}_{(k)} {\overset{\rm a.s.}{=}} \im  L^1_{(k)} + \im  L^2_{(k)} $.
	This implies that (\ref{eq:L^1_and_L^2_rank}) can be rewritten as
	\begin{align*}
		\rank Z^{(q,T)}_{(k)} {\overset{\rm a.s.}{=}} \rank L^1_{(k)} \!+\! \rank L^2_{(k)} \!-\! \dim(\im \! L^1_{(k)}\!\cap\! \im\!  L^2_{(k)}).
	\end{align*}
	
	We then show that $ \im L^1_{(k)} \cap \im L^2_{(k)} = \{0\} $.
	To this end, assume for the sake of contradiction that $ \im L^1_{(k)} \cap \im L^2_{(k)} \neq \{0\} $. Then, there exists a nonzero vector such that $ v \in \im L_{(k)}^1 $ and $ v \in \im L_{(k)}^2 $, which implies
	\begin{align*}
		\left[
		\begin{array}{c}
			I_{mq} \\ \mathcal{T}^q
		\end{array}\right] u =
		\left[
		\begin{array}{c}
			0 \\ \mathcal{O}^q
		\end{array}\right]x,~~~  \exists u \in \R^{mq}, x \in \R^{n}.
	\end{align*}
	This yields $ u = 0 $, which contradicts the assumption that $ v $ is nonzero, and thus $ \im L^1_{(k)} \cap \im L^2_{(k)} = \{0\} $.
	Hence, we derive
	\begin{align}
		\rank Z^{(q,T)}_{(k)} {\overset{\rm a.s.}{=}} \rank L^1_{(k)} + \rank L^2_{(k)}.
	\end{align}
	
	{From Lemma~\ref{lemma:U_k}, $  U^{(q,T)}_{(k)} $ has full row rank for all $ k \in [0,N-T-q+1]$ almost surely, which implies
		\begin{align*}
			\rank L^1_{(k)} \overset{\mathrm{a.s.}}{=} mq,~\forall k \in [0,N-T-q+1].
		\end{align*}
		Also, from Lemma~\ref{lemma:X_rank}, $ \rank X^{(T)}_{(k)} $ has full row rank for all $ k \in [0,N-T+1]$ almost surely, which implies
		\begin{align*}
			\rank L^2_{(k)} \!=\! \rank \cO^q X^{(T)}_{(k)} \!\overset{\mathrm{a.s.}}{=} \!\rank \cO^q,~\forall k \!\in\! [0,N\!-\!T\!+\!1].
		\end{align*}
		Therefore, we have (\ref{eq:lemma_Z_rank}).
		Specifically, if $ q \geq n $, $ \cO^q $ has full column rank because of the system observability, which yields (\ref{eq:lemma_Z_rank_n}).
	}
\end{proof}

{One can observe from this proposition that, if the attack does not exist over the entire data and $ T \geq mq + n + 1  $, then the rank of $ Z^{(q,T)}_{(k)} $ is constant for all $ k $ almost surely.}

{
We then present the following proposition on the rank of $ Z^{(q,T)}_{(k)} $ in the presence of sensor attacks.}
\begin{proposition}
\label{proposition:detection_attack}
{Suppose that Assumptions \ref{assumption:system}--\ref{assumption:clear_data} hold, i.e., the system is controllable and observable, $ A,B,C $ are unknown, the I/O data with a partially attack-free interval are given, $ u(k) $ is designed by (\ref{eq:control}), and $ |{\mathcal{A}^*}| = \ell  $.}
If $ q \in [n+1, N-T+1] $, $ T \in [mq+pq, N-q+1] $, and $ \tau \geq T+q-1 $, then
\begin{align}
	\label{eq:Z_condition_clear}
	\hspace{-5mm}\rank Z^{(q,T)}_{(k)} \! \left\lbrace
	\begin{array}{ll}
		\!\!  {\overset{\mathrm{a.s.}}{=}}  mq \!+\! n, & \!\!\forall k \!\in \![k_0, k_0 \!+\! \tau \!-\!T \!-\! q \!+\! 1],  \\
		\!\!  {\overset{\mathrm{a.s.}}{\neq}}  mq \!+\! n, & \!\!\exists k \!\notin \![k_0, k_0 \!+\! \tau \!-\!T \!-\! q \!+\! 1].
	\end{array}
	\right.\!\!\!\!
\end{align}
\end{proposition}
\vspace{1mm}
\begin{proof}
To improve readability, for each $ k $, we define two matrices $ L_{(k)} $ and $ M_{(k)} $ as follows:
\begin{align*}
	Z^{(q,T)}_{(k)} \!=\! \left[\!
	\begin{array}{c}
		U^{(q,T)}_{(k)}\\ Y^{(q,T)}_{(k)}
	\end{array}\!\right] \!=\! \underbrace{\left[\!
		\begin{array}{c}
			U^{(q,T)}_{(k)} \\ \mathcal{T}^q U^{(q,T)}_{(k)} \!+\! \mathcal{O}^q  X^{(T)}_{(k)}
		\end{array}\!\right]}_{L_{(k)}} \! + \!\underbrace{\left[\!
		\begin{array}{c}
			0 \\ \Lambda^{(q,T)}_{(k)}
		\end{array}\!\right]}_{M_{(k)}}.
\end{align*}
{Since $ q \in [n+1, N-T+1] $ and $ T \in [mq+pq, N-q+1] $, the condition $ T \geq mq + n + 1 $ is met, and thus, from  Proposition~\ref{proposition:detection_attack-free}, we have
	\begin{align}
		\label{eq:rank_L_k}
		\rank L_{(k)} \overset{\mathrm{a.s.}}{=} mq \!+\! n,\!~\forall k \!\in\! [0,N\!-\!T\!-\!q\!+\!1].
\end{align}}
From the clean interval definition (cf. Fig.~\ref{fig:Relationship_Lambda}), it follows that $ \Lambda^{(q,T)}_{(k)} = 0 $ for all $ k \in [k_0, k_0 + \tau-T - q + 1] $, which implies
\begin{align*}
	\rank Z^{(q,T)}_{(k)} \!\!=\! \rank L_{(k)} {\overset{\mathrm{a.s.}}{=}}mq \!+\! n,\!~\forall k \!\in\! [k_0, k_0 \!+\! \tau \!-\!T \!-\! q \!+\! 1],
\end{align*}
{which indicates that the first statement in (\ref{eq:Z_condition_clear}) holds.}

We next show that there exists $ k \notin [k_0, k_0 + \tau-T - q + 1] $ such that $ \rank Z^{(q,T)}_{(k)} {\overset{\mathrm{a.s.}}{\neq}} mq+n  $.
Without loss of generality, assume that there exists a transition interval after the partially attack-free interval $ \K_0 $ and define $ k^* \triangleq k_0 + \tau-T - q + 2 $, which is the first time instant in the transition interval after $ \K_0 $ in the matrix sense\footnote{{In this proof, we assume that there exists a transition interval after $ \K_0 $.
		If there is no transition interval after $ \K_0 $ (i.e., $ k_0 + \tau -1 = N-1 $), one can instead analyze using the time instant \textit{before} $ \K_0 $. If no such transition exists on either side, then the whole dataset is attack-free and the rank remains constant as shown in Proposition~\ref{proposition:detection_attack-free}.}}.
Then, since now $ T \leq \tau - q+1$, the attack Hankel matrix at time $ k^* $ has the following structure:
\begin{align}
	\label{eq:a_observability_0}
	\Lambda_{(k^*)}^{(q,T)} = \left[
	\begin{array}{ccccc}
		0 & \cdots & 0 & 0 \\
		\vdots & \ddots & \vdots & \vdots \\
		0 & \cdots & 0 & 0 \\
		0 & \cdots & 0 & a(k_0 + \tau)
	\end{array}\right],
\end{align}
{and thus $ M_{(k^*)} $ has only one nonzero column in the last column.
	Let $ e_T \in \R^T $ be the $ T $th standard basis vector and define
	\begin{align*}
		\delta \triangleq M_{(k^*)} e_T = \left[
		\begin{array}{cccc}
			0 & \cdots & 0 & a(k_0 + \tau)^\top
		\end{array}\right]^\top\in\R^{mq+pq}.
	\end{align*}
	Then, we have $ \im Z^{(q,T)}_{(k^*)}  = \im L_{(k^*)} + \mathrm{span}\{\delta\} $.}

{
	We next claim that $ \delta \notin \im L_{(k^*)} $.
	Suppose for contradiction that $ \delta \in \im L_{(k^*)} $.
	Then, there exists $ \alpha \in \R^T $ such that $ L_{(k^*)}\alpha = \delta $.
	Comparing the first $ mq $ rows yields $ U^{(q,T)}_{(k^*)} \alpha = 0 $, and hence it holds that $ \mathcal{T}^q U^{(q,T)}_{(k^*)} \alpha = 0 $.
	Therefore, by comparing the last $ pq $ rows, we obtain $ \cO^q X^{(T)}_{(k^*)}\alpha = \delta_y $, where $ \delta_y $ denotes the last $ pq $ entries of $ \delta $.}
Let $ x \triangleq  X^{(T)}_{(k^*)}\alpha \in \R^n $.
Then, from the construction of $ \mathcal{O}^q $ and {$ \delta_y $}, we have $ \mathcal{O}^{q-1} x = 0 $ and $ CA^{q-1}x = a(k_0 + \tau) $.
Recalling that $ q \geq n+1 $ and the system is observable, $ \mathcal{O}^{q-1} $ has full column rank and $ \ker \mathcal{O}^{q-1} = \{0\} $, which implies $ x = 0 $, {and hence, $ CA^{q-1}x = 0 $.
	This contradicts $ a(k_0 + \tau) \neq 0 $, and thus, $ \delta \notin \im L_{(k^*)} $.}

{
	By using (\ref{eq:Grassmann_formula}) again, we have
	\begin{align*}
		\rank Z^{(q,T)}_{(k^*)} \!=\! \rank L_{(k^*)} \!+\! \rank \delta\!-\!\dim(\im L_{(k^*)} \cap \mathrm{span}\{\delta\}).
	\end{align*}
	From (\ref{eq:rank_L_k}), $ \rank L_{(k^*)} \overset{\rm a.s.}{=} mq+n $.
	By construction, $ \rank \delta = 1 $.
	Additionally,  $ \delta \notin \im L_{(k^*)} $ implies $ \dim(\im L_{(k^*)} \cap \mathrm{span}\{\delta\}) = 0 $.
	Therefore, $ \rank Z^{(q,T)}_{(k^*)}  \overset{\rm a.s.}{=}mq+n+1 $, which proves the second statement in (\ref{eq:Z_condition_clear}).}
\end{proof}

This proposition states that the rank of $  Z^{(q,T)}_{(k)} $ cannot be kept constant through all intervals {almost surely} under the conditions of Proposition~\ref{proposition:detection_attack}.
{Specifically, under these conditions, the rank of $  Z^{(q,T)}_{(k)} $ inevitably deviates during the transition interval.
Note that the conditions of Proposition~\ref{proposition:detection_attack} are sufficient for Proposition~\ref{proposition:detection_attack-free}, i.e., if $ T \geq mq+pq $ and $ q \geq n+1 $, then $ T \geq mq + n+1 $.}

{Combining the results of Propositions 1 and 2, under the conditions of Proposition~\ref{proposition:detection_attack}, we can detect malicious sensor attacks in the data by checking the rank of $ Z_{(k)}^{(q, T)} $ for all $k$ and observing the rank variations: if the rank is constant through all intervals, we conclude that there is no attack in the data; otherwise, some data are compromised.}
Note that this detection scheme does not restrict the number of sensor attacks $ \ell $.
Thus, even if $ \ell = p $, i.e., all sensor outputs are compromised except for the attack-free interval, the attack can be detected as long as the conditions of Proposition~\ref{proposition:detection_attack} hold.
Focusing on the length of the partially attack-free interval $ \tau $, if $ \tau $ is smaller than $ T+q-1 $ (i.e., the attack-free interval is too short), one may not observe the rank variations and detect the attacks.
Hence, as mentioned in Remark~\ref{remark:clean_data}, it is recommended to use a large amount of data to enhance the likelihood of a long-term attack-free interval existing.

{
In practice, however, the conditions in Proposition~\ref{proposition:detection_attack} depend on the unknown system order $ n $, and hence, it is in general impossible to determine the appropriate order $ q $.
To obtain a practical procedure that does not require prior knowledge of $ n $, we propose a heuristic data-driven detector that evaluates the rank profile in a test window.}

\subsection{Attack Detection Algorithm}

\begin{algorithm}[t]
\caption{Data-driven attack detection with partially attack-free data}
\label{algorithm:detection_clean}
\begin{algorithmic}[1]
	\Statex \hspace{-6mm} \textbf{Input:} $ u^{[N-1]} $, $ y^{[N-1]} $, $ m $, $p$, $N$, $ {L} $, and {$ q_{\max} $}
	\Statex \hspace{-7mm} \textbf{Output:} {``Attack Detected'' or ``No-Attack Detected''}
	\For{{$ q = q_{\max}\!-\!L\!+\!1,q_{\max}\!-\!L\!+\!2,\ldots,q_{\max} $}}
	\State Set $ T $ as $ T = mq + pq $.
	\For{all $  k \in [0,N-T-q+1] $}
	\State {Construct $ Z^{(q,T)}_{(k)} $ based on (\ref{eq:Z_matrix_k}).}
	\State Compute $ r^q_{(k)} = \rank Z^{(q,T)}_{(k)} $.
	\EndFor
	\If{$r^q_{(k)} $ is not {constant} over $  k $}
	\State \textbf{return} {``Attack Detected''}
	\EndIf
	\EndFor
	\State \textbf{return} {``No-Attack Detected''}
\end{algorithmic}
\end{algorithm}

{
The proposed heuristic data-driven attack detection algorithm is summarized in Algorithm~\ref{algorithm:detection_clean}.
Since the system order $ n $ is unknown, the algorithm scans $r^q_{(k)} \triangleq \rank  Z_{(k)}^{(q, T)} $ over a range of window sizes $ q $.
To mitigate the ambiguous regime $ q < n+1 $, the decision is made by testing the window $ q \in [q_{\max}-L+1,q_{\max}] $, where $ q_{\max} $ and $ L $ are integers to be designed.
The detector declares an attack if a rank variation is observed for any tested $ q $.
According to Proposition~\ref{proposition:detection_attack}, the algorithm returns ``Attack Detected'' almost surely if the tested set contains at least one $ q \geq n+1 $ and the data include an attack-free interval of length $ \tau \geq T+q-1 $.
If a rank variation is not observed in the tested set, the algorithm returns ``No-Attack Detected''.
Note that this should not be interpreted as a certificate of absence of attacks.
In particular, attacks may remain undetectable if the attack-free interval is short enough or $ q \geq n+1 $ cannot be realized in the tested window sizes.
}

{
When the upper bound of the system order, denoted by $ \bar n $, is available, it is reasonable to execute Algorithm~\ref{algorithm:detection_clean} by setting $ q = \bar n + 1 $.
In this case, the algorithm returns ``Attack Detected'' almost surely if the attack-free interval satisfies $ \tau \geq T + q -1 $.}

{
We next discuss the computational complexity of this algorithm.
Since the matrix $ Z_{(k)}^{(q, T)} $ is square due to $ T = mq + pq $, the cost of the rank computation in Line 5 is $ O((mq+pq)^3) $.
Since the rank condition is computed for all $ k \in [0,N-T-q+1] $ and $  q \in [q_{\max}-L+1,q_{\max}] $, the total complexity is
\begin{align*}
	\sum_{q=q_{\max}-L+1}^{q_{\max}} O\left(\left(N-(m+p+1)q\right)\cdot \left((mq+pq)^3\right)\right),
\end{align*}
which is bounded by
\begin{align}
	O\left(LN(mq_{\max}+pq_{\max})^3\right).
\end{align}
This highlights $q_{\max}$ as the primary computational knob, with cubic scaling, whereas $ L$ and $N$ enter linearly.
With the choice $ T = mq + pq $, the sufficient condition in Proposition~\ref{proposition:detection_attack} requires an attack-free interval of length $ \tau \geq T + q- 1 = (m+p+1)q - 1 $.
To guarantee detectability under Proposition~\ref{proposition:detection_attack}, larger $ q_{\max} $ increases the chance that some tested $ q $ satisfies the unknown condition $ q \geq n+1 $, but it also increases the required attack-free length $ \tau \geq (m+p+1)q - 1 $ and the computational burden.
In practice, if a large amount of data is obtained, it is recommended to choose $ q_{\max} $ as large as permitted by the data length and computational budget, and then test $ L $ to increase the likelihood that at least one tested $ q $ satisfies $ q \geq n+1 $.
Since increasing $ L $ enlarges the set of tested window sizes, using a relatively small $ L $ is reasonable.}

\section{Data-Driven Attack Identification with Partially Attack-free Data}
\label{section:identification}
We next consider Problem~\ref{problem:sensor_identifier}, i.e., the data-driven attack identification problem with partially attack-free data.

\subsection{Attack Identification Condition}
The following theorem provides the attack identification condition with partially attack-free data.
\begin{theorem}
\label{theorem:clean_attack_identification}
{Suppose that Assumptions \ref{assumption:system}--\ref{assumption:clear_data} hold.
	Also, assume that $ q \in [n+1, N-T+1] $, $ T \in [m(q+n)+n, N-q+1] $, and $ \tau \geq T+q-1 $.}
For any $ t \in [0,N-T-q+1] $, define $ K^q_{(t)}\triangleq (U^2_{(t)})^\top \in\R^{(mq+pq-r^q_{(t)})\times (mq+pq)} $, where $ U^2_{(t)} $ is obtained by the SVD of
\begin{align}
	\label{eq:SVD}
	Z^{(q,T)}_{(t)} \!\!=\! \left[
	\begin{array}{cc}
		\!\!U^1_{(t)}\! &\! U^2_{(t)}\!\!
	\end{array}\right]\left[
	\begin{array}{cc}
		\!\Sigma^1_{(t)}\! & 0 \\ 0 & \!\Sigma^2_{(t)}(\approx\! 0) \!
	\end{array}\right]\left[
	\begin{array}{c}
		\!\!(V^1_{(t)})^\top \!\!\\\!\! (V^2_{(t)})^\top\!\!
	\end{array}\right]
\end{align}
and $ r^q_{(t)} =  \rank Z^{(q,T)}_{(t)} $.
Also, for $ t \in [0,N-T-q+1]  $, $ k \in [0, N-q] $, and $ \Gamma \subseteq \cP $, define
\begin{align}
	\label{eq:gamma}
	\gamma^\Gamma_{(t, k)}  \triangleq P^\Gamma_{(t)} K^q_{(t)} z^{[k,k+q-1]},
\end{align}
where $ P^\Gamma_{(t)} $ is a filter matrix whose rows form an orthonormal basis for the left null space of $ Q^2_{(t)} \mathbb{I}^\Gamma_q  $, i.e.,
\begin{align}
	\label{eq:P^Gamma}
	P_{(t)}^\Gamma Q^2_{(t)}   \mathbb{I}^\Gamma_q  = 0,
\end{align}
where $ Q^2_{(t)} \in\R^{(mq+pq-r^q_{(t)})\times pq}  $ can be obtained from $ K^q_{(t)} $ as
\begin{align}
	K^q_{(t)} = \left[
	\begin{array}{cc}
		Q^1_{(t)} & Q^2_{(t)}
	\end{array}\right],
\end{align}
and
\begin{align}
	\label{eq:I^Gamma_q}
	\mathbb{I}^\Gamma_q \triangleq \mathrm{blockdiag}(\underbrace{I^\Gamma_p,\ldots,I^\Gamma_p}_{q~\mathrm{times}}) \in \R^{pq \times |\Gamma|q},
\end{align}
where $ I^{\Gamma}_p \in\R^{p\times |\Gamma|}$ is the submatrix of $ I_p $ with the columns in $ \Gamma $.
Then, {for all $ \Gamma \subseteq \cP $,
	\begin{enumerate}
		\item If $ \A^* \subseteq \Gamma $, then
		\begin{align}
			\label{eq:theorem_first_eq}
			\hspace{-5mm}\gamma^\Gamma_{(t, k)} \! \overset{\mathrm{a.s.}}{=} \!0,~\forall k \!\in\! [0,N\!-\!q],~\forall t \!\in\! [0,N\!-\!T\!-\!q\!+\!1].
		\end{align}
		\item If $ \A^* \nsubseteq \Gamma $, then
		\begin{align}
			\label{eq:theorem_second_eq}
			\hspace{-7mm}\exists k \!\in \![0,N\!-\!q],~\exists t \!\in \![0,N\!-\!T\!-\!q\!+\!1]~\mathrm{s.t.}~\gamma^\Gamma_{(t, k)} \! \overset{\mathrm{a.s.}}{\neq} \!0.
		\end{align}
\end{enumerate}}
\vspace{0.1mm}
\end{theorem}
\begin{proof}
See Appendix~\ref{appendix:proof_of_theorem}.
\end{proof}

Under the conditions of Theorem~\ref{theorem:clean_attack_identification}, for each $ \Gamma \subseteq \cP $, if the compromised sensor set $ {\A^*} $ satisfies $ {\A^*} \subseteq \Gamma $, then $ \gamma^\Gamma_{(t, k)} \overset{\mathrm{a.s.}}{=} 0 $ for the two time indices $ k \in [0, N-q]$ and $ t \in [0,N-T-q+1]  $.
On the other hand, if $ {\A^*} \nsubseteq \Gamma $, then there exist $ t $ and $ k $ such that $ \gamma^\Gamma_{(t, k)} \overset{\mathrm{a.s.}}{\neq} 0 $.
Accordingly, for each $ \Gamma \subseteq \cP $, by computing $ \gamma^\Gamma_{(t, k)} $ for all $ k\in[0,N-q] $ and $ t \in [0,N-T-q+1] $, and observing the variation in its values, we can identify the compromised sensor set using the I/O data. The concrete algorithm is discussed in the next subsection.
{Note that $ \gamma^\Gamma_{(t, k)}  $ depends on two time indices, $ k $ and $ t $.
The index $ t $ is used to construct the matrix $ Z^{(q,T)}_{(t)} $ and its associated matrices $ K^q_{(t)} $ and $ P^\Gamma_{(t)} $, while $ k $ is used to create the I/O vector $ z^{[k,k+q-1]} $.}

\begin{remark}
\label{remark:input}
{
	The additive watermark in Assumption~\ref{assumption:input} provides persistent excitation in a probabilistic sense and thereby enforces the rank conditions (Lemmas \ref{lemma:U_k}--\ref{lemma:U-X_rank}) with probability one, which enables Propositions \ref{proposition:detection_attack-free}--\ref{proposition:detection_attack} and Theorem~\ref{theorem:clean_attack_identification} to hold almost surely.
	Without watermarking (i.e., $ \varphi = 0 $), the same guarantees would require an explicit design of $ u_{\rm nom} $ to achieve the persistent excitation in each interval window and may fail in closed-loop operation. }
\end{remark}

\subsection{Attack Identification Algorithm}

\begin{algorithm}[t]
\caption{Data-driven attack identification with partially attack-free data}
\label{algorithm:identification_clean}
\begin{algorithmic}[1]
	\Statex \hspace{-6mm} \textbf{Input:} $ u^{[N-1]} $, $ y^{[N-1]} $, $ m $, $p$, $N$, and $ {q=q_{\max}} $
	\Statex \hspace{-6mm} \textbf{Output:} Compromised sensor set $ {\mathcal{A}^*} $
	\State Set $ T $ as $ {T = m(2q-1) + q-1} $.
	\For{all $ t \in [0,N-T -q + 1]$}
	\State Compute the SVD of (\ref{eq:SVD}) and obtain $ K^q_{(t)} $.
	\For{each $ \Gamma \subseteq \cP $}
	\State Compute $ P^\Gamma_{(t)} $ based on Eqs. (\ref{eq:P^Gamma})--(\ref{eq:I^Gamma_q}).
	\For{all $ k \in [0,N-q] $}
	\State Compute $ \gamma^\Gamma_{(t, k)} $ based on (\ref{eq:gamma}).
	\EndFor
	\EndFor
	\EndFor
	\State Collect all possible sets $ \Gamma $ such that $ \gamma^\Gamma_{(t, k)} \! = \! 0,~\!\forall k,t $ into a set $ \mathcal{J}$.
	\State \textbf{return} $ {\mathcal{A}^*} = \arg \min_{\Gamma \in \mathcal{J}}~|\Gamma| $
\end{algorithmic}
\end{algorithm}

{Based on Theorem~\ref{theorem:clean_attack_identification}, the heuristic algorithm for attack identification using partially attack-free data is given as Algorithm~\ref{algorithm:identification_clean}.
This algorithm is executed after Algorithm~\ref{algorithm:detection_clean} returns ``Attack Detected'', and is valid under the conditions of Theorem~\ref{theorem:clean_attack_identification}.
As in the previous section, since the system order $ n $ is unknown, it is reasonable to set $ q = q_{\max} $ based on the parameter $ q_{\max} $ used in Algorithm~\ref{algorithm:detection_clean} to avoid the ambiguous scenario $ q < n+1 $.
If $ q = q_{\max} \geq n+1 $, then $ T $, which is set in Line 1 of Algorithm~\ref{algorithm:identification_clean}, satisfies the condition of Theorem~\ref{theorem:clean_attack_identification}, i.e., $ T \geq m(q+n)+n $.
In Lines 2--10, $ \gamma^\Gamma_{(t, k)} $ is computed for all $ k\in[0,N-q]$, $ t \in [0,N-T-q+1] $, and subsets $ \Gamma \subseteq \cP $.}
To satisfy (\ref{eq:P^Gamma}), one can choose $ P^\Gamma_{(t)} $ whose rows form an orthonormal basis for the left null space of $ Q^2_{(t)} \mathbb{I}^\Gamma_q  $.
This basis can be computed, for example, via SVD or QR decomposition.
{In Line 11, construct the set $ \mathcal{J} $ by collecting all possible sets $ \Gamma $ such that $ \gamma^\Gamma_{(t, k)}  =  0 $ for all $ k $ and $ t $, namely,
\begin{align*}
	\J \triangleq \{\Gamma \subseteq \cP:\gamma^\Gamma_{(t, k)} = 0,~\forall k,t \}.
\end{align*}
From Theorem~\ref{theorem:clean_attack_identification}, if $ \A^* \subseteq \Gamma $, then $ \Gamma \in \mathcal{J} $ almost surely, i.e., we have $ \J \overset{\rm a.s.}{=} \{\Gamma \subseteq \cP : \A^* \subseteq \Gamma \} $.
The unique minimum-cardinality set in this family is $ \A^* $, because any strict superset of $ \A^* $ has strictly larger cardinality.
Therefore, in Line 12, selecting $ \arg \min_{\Gamma \in \mathcal{J}}~|\Gamma| $ recovers $ \A^* $ uniquely if the conditions of Theorem~\ref{theorem:clean_attack_identification} are satisfied.}
As with the detection algorithm, note that this identification algorithm does not impose a limit on the number of compromised sensors.
If all sensors are compromised, i.e., $ \A^* = \cP $, it holds that $ \gamma^\Gamma_{(t, k)}  {\overset{\rm a.s.}{=}} 0 $ for all $ t, k $ only when $ \Gamma = \cP $, under the conditions of Theorem~\ref{theorem:clean_attack_identification}.

{
We next discuss the computational complexity of this algorithm.
For Line 3, based on traditional algorithms to compute SVD (e.g., \cite[Chapter 45]{SVD}), the computational complexity of the SVD of each $ Z^{(q,T)}_{(t)} $ is
\begin{align*}
o_{svd} \triangleq O\left((mq+pq)\cdot T \cdot \min(mq+pq, T)\right).
\end{align*}
Similarly, for Line 5, the SVD-based computation of each $ P^\Gamma_{(t)} $ requires
\begin{align*}
o_{p} \triangleq O\left((mq\!+\!pq\!-\!r^q_{(t)})\cdot(|\Gamma|q)\cdot\min(mq\!+\!pq\!-\!r^q_{(t)}, |\Gamma|q) \right)\!.
\end{align*}
Further, for Line 7, the computation of each $ \gamma^\Gamma_{(t,k)} $ requires
\begin{align*}
o_{\gamma} \triangleq O\left((mq+pq-r^q_{(t)})\cdot(mq+pq)\right).
\end{align*}
Since the index $ t $ is iterated $ N_t \triangleq N-T-q+1 $ times and $ k $ is iterated $ N_k \triangleq N-q+1 $ times, the total complexity is given by
\begin{align}
O\left(N_t o_{svd}+N_t \sum_{\Gamma \subseteq \cP}\left(o_p + N_k o_\gamma\right)\right),
\end{align}
which is bounded by
\begin{align}
O\left(N_t o_{svd}+N_t \cdot 2^p \cdot \left(o_p + N_k o_\gamma\right)\right).
\end{align}
This bound highlights the exponential dependence on the number of sensors $ p $, and thus Algorithm~\ref{algorithm:identification_clean} is computationally intensive.
Reducing the computational burden is an important future direction.}


\begin{remark}
\label{remark:computation}
In computational implementation, the residual $ \gamma_{(t,k)}^\Gamma $ is generally not exactly zero even for $ {\A^*} \subseteq \Gamma $, due to numerical errors.
It is therefore reasonable to introduce a small threshold $ \epsilon > 0 $ and regard $ \gamma_{(t,k)}^\Gamma $ as zero if $ \| \gamma_{(t,k)}^\Gamma \|_2 \leq \epsilon $.
The choice of $ \epsilon $ should be based on the expected numerical precision.
{
Likewise, the rank test in Algorithm \ref{algorithm:detection_clean} is carried out using a numerical rank (e.g., via SVD), where singular values below a tolerance are treated as zero.
Note that Algorithm~\ref{algorithm:identification_clean} already computes the SVD of (\ref{eq:SVD}) in Line 3, and thus $ r^q_{(t)} =  \rank Z^{(q,T)}_{(t)} $ is obtained with essentially no additional cost using a SVD-based numerical rank.}
\end{remark}

\section{Numerical Simulations}
\label{section:simulation}
In this section, we demonstrate the effectiveness of the proposed data-driven framework for attack detection and identification through numerical simulations.
{Specifically, we randomly sample the system matrices and confirm empirical detection and identification rates for several attack-free interval lengths.}
We compare the proposed framework {with the method presented in \cite{2021L-CSSPasquialetti}} and robust principal component analysis (ROBPCA).

\subsection{Simulation Setting}
{
Consider a linear time-invariant system (\ref{eq:sytem_model}) of state-space dimension $ n = 20 $ defined below:
\begin{align*}
A = \left[
\begin{array}{c|ccc}
	0 & 1 & & \\
	\vdots & & \ddots & \\
	0 & & & 1 \\ \hline
	1 & 1 & \cdots & 1
\end{array}\right] \in \R^{20 \times 20},
\end{align*}
with $ m = 1 $ actuator and $ p = 5 $ sensors.
In each trial, $ B \in \R^{20} $ is chosen at random from the canonical basis of $ \R^n $ and $ C \in \R^{5 \times 20} $ has i.i.d. Gaussian entries.
We assume that $ N=1000 $ data samples are obtained.
For all trials, the input is defined as $ u(k) = 0.1 \sin(0.01 k) + w(k) $, where $ w(k) $ is an i.i.d. Gaussian watermark with $ \varphi^2 = 10^{-8} $, independent of the nominal input.
We vary the attack-free interval length $ \tau $ and, for each $ \tau $, run 30 independent trials with newly sampled $ B, C $, watermark realizations, and initial conditions.
Assume that the compromised sensor set is given as $ \A^* =\{1,2,3,4\} $, i.e., all sensors except the last one are compromised outside the attack-free interval.
In this simulation, we consider two attack scenarios:
The first scenario adopts the following simple additive attack:
\begin{align}
\label{eq:first_attack}
a(k) \!=\! \left[0.1,0.1,0.1,0.1,0\right]^\top\!\!,~\forall k \!\in\! [0,N\!-\!1]\setminus\K_0.
\end{align}
In the second scenario, we assume that the attacker knows the model and designs the following sophisticated attack based on the knowledge:
\begin{align}
\label{eq:second_attack}
a_i(k) \!=\! -2\left(C_i x(k) \!+\! C_i x(\tau)\right),~\!\forall k \!\in\! [0,N\!-\!1]\setminus\K_0,
\end{align}
for $ i \in \A^*=\{1,2,3,4\} $ and $ a_5(k) \equiv 0 $ for all $ k $.}

{
For the proposed framework, we follow Algorithms \ref{algorithm:detection_clean}--\ref{algorithm:identification_clean} with $ q_{\max} = 25 $ and $ L = 3 $ and compute ranks numerically using an SVD-based tolerance $ 10^{-15} $; similarly, we apply a small threshold $ \epsilon = 10^{-10} $ to decide whether quantities that are theoretically zero are treated as zero (cf. Remark~\ref{remark:computation}).}

\begin{figure*}[t]
\begin{center}
\includegraphics[width=\linewidth]{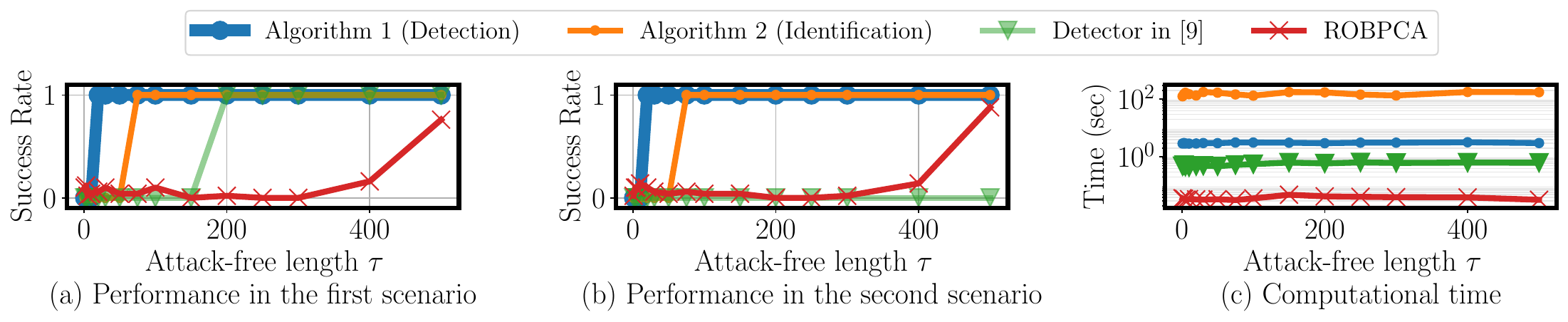}
\vspace{-7mm}
\caption{Detection/identification performances and computational performances of the proposed algorithms, detector in \cite{2021L-CSSPasquialetti}, and ROBPCA.}
\vspace{-7mm}
\label{fig:simulation_result}
\end{center}
\end{figure*}

\subsection{Benchmarks}
As benchmarks, we compare against {the data-driven detector of \cite{2021L-CSSPasquialetti}} and the ROBPCA \cite{ROBPCA-01,ROBPCA-03,ROBPCA-04} method\footnote{In this work, we use the Robpy Python package \cite{Robpy} to implement the ROBPCA method and compute the orthogonal and score distances.}.
The ROBPCA method combines ideas of both projection pursuit and robust covariance estimation, and uses two metrics to distinguish anomalies in data: orthogonal distance and score distance.
To classify the data, cutoff values on these distances are determined using the corresponding distribution to have an exceeding probability of 2.5\% \cite{ROBPCA-01}.
Regular data have small orthogonal and score distances and do not exceed the cutoffs.
Conversely, if some data exceed the orthogonal or score distance cutoff, the data may contain anomalies.
This method is well-suited for analyzing high-dimensional data and has been applied to anomaly detection and outlier diagnosis.
In this simulation, we apply ROBPCA to the compromised output data $ y^{[N-1]} $.

{
For the detector of \cite{2021L-CSSPasquialetti}, we set the detection threshold as $ 0.1 $ and the window size for a Hankel matrix to $ 166 $, which follows the paper's heuristic algorithm.
For details, see \cite[Remark 2]{2021L-CSSPasquialetti}.
The data-driven algorithm of \cite{2021L-CSSPasquialetti} requires an initial attack-free segment, and thus, for a fair comparison, we place the clean interval at the beginning of the dataset (i.e., $ k_0 = 0 $) in all experiments.
In contrast, note that our method does not require an initial clean segment; attack detection and identification remain possible as long as an attack-free interval of sufficient length exists anywhere in the data record.}

\subsection{Simulation Results}
{
Figure~\ref{fig:simulation_result} shows the attack detection/identification performances and computational performances\footnote{The simulations are performed on a laptop equipped with an Intel Core i7-1165G7 2.80GHz and 16~GB of memory.} of Algorithms \ref{algorithm:detection_clean}--\ref{algorithm:identification_clean}, the data-driven detector \cite{2021L-CSSPasquialetti}, and ROBPCA, where Fig.~\ref{fig:simulation_result}-(a) illustrates the detection/identification performances in the first attack scenario of (\ref{eq:first_attack}), Fig.~\ref{fig:simulation_result}-(b) illustrates the detection/identification performances in the second attack scenario of (\ref{eq:second_attack}), and Fig.~\ref{fig:simulation_result}-(c) illustrates the average computational time in the first attack scenario.
For each $\tau$, the success rates are computed over 30 independent Monte Carlo trials.
For each method, a trial is counted as a success under the following conditions:
Algorithm~\ref{algorithm:detection_clean} returns "Attack Detected".
Algorithm~\ref{algorithm:identification_clean} exactly recovers the true compromised sensor set $ \A^* $.
The detection monitor in the detector of \cite{2021L-CSSPasquialetti} exceeds the designed threshold.
At least one dataset exceeds the orthogonal or score distance cutoff in the ROBPCA.
}

{
From Figs.~\ref{fig:simulation_result}-(a) and \ref{fig:simulation_result}-(b), we observe that the proposed detector (Algorithm~\ref{algorithm:detection_clean}) achieves successful detection if $ \tau \geq 20 $.
The proposed identification algorithm (Algorithm~\ref{algorithm:identification_clean}) requires a slightly longer attack-free segment than Algorithm~\ref{algorithm:detection_clean}, but it also rapidly reaches a successful identification after $ \tau \geq 75 $.
The proposed algorithms remain effective against the sophisticated attack, even when $ \ell = 4 $ out of $ p = 5 $ sensors are compromised.
In contrast, the detector in \cite{2021L-CSSPasquialetti} performs well against the simple additive attack only when a sufficiently long attack-free interval is placed at the beginning of the dataset, but it fails to detect the model-aware attack, even for large $\tau$.
ROBPCA improves only when $\tau$ becomes very large, consistent with its reliance on having a dominant fraction of nominal samples.
Note that, in the present simulations, Algorithms \ref{algorithm:detection_clean}--\ref{algorithm:identification_clean} succeed even for values of $ \tau $ that are smaller than the sufficient bounds of Proposition 2 and Theorem 1, but this does not contradict the theory.
When $\tau < T+q-1$, the theory simply does not provide a worst-case guarantee, but detection/identification can still succeed if the attack does not satisfy the algebraic constraints required to keep the rank profile and the left-kernel residuals consistent across all windows.}

{
For the computational performance depicted in Fig.~\ref{fig:simulation_result}-(c), the detection algorithm (Algorithm~\ref{algorithm:detection_clean}) is relatively fast, whereas the identification algorithm (Algorithm~\ref{algorithm:identification_clean}) is substantially more expensive due to the exhaustive search over sensor subsets, consistent with the complexity analysis in the previous section.
The benchmark methods are faster, but this comes at the cost of markedly inferior attack-detection performance.}

\section{Conclusion}
\label{section:conclusion}
This paper studied data-driven attack detection and identification in a model-free setting.
{We considered a scenario in which the available data may be compromised, but contain an unknown, attack-free interval.
Under the assumption that the control input contains a small stochastic watermark, we established sufficient conditions for data-driven attack detection and identification.
Also, we developed data-driven algorithms and characterized their computational complexity.}
Through numerical simulations, we demonstrated the efficacy of the proposed methods.

Several directions remain for future work.
One direction is to develop attack detection and identification schemes that explicitly account for process and measurement noise.
Identifying the clean interval from partially attack-free datasets is another interesting direction.
\appendices

\section{Proof of Lemma~\ref{lemma:U_k}}
\label{appendix:proof_lemma_U}
{For notational simplicity, for fixed $ k \in [0,N-T-q+1] $, denote the input data from time $ k $ to $ k + T+ q- 2 $ by $ \xi \triangleq u^{[k,k+T+q-2]} $.
By construction, $ U^{(q,T)}_{(k)} = \mathscr{H}_q(\xi) $, and hence $ U^{(q,T)}_{(k)} $ is an affine function of $ \xi $.
Let $ \{M_j(\xi)\}_{j=1}^J $ denote the collection of all $ mq \times mq $ submatrices of $ U^{(q,T)}_{(k)} $, where $ J \triangleq \binom{T}{mq}. $
Also, define the scalar function based on the matrix determinant as $ g(\xi) \triangleq \sum_{j=1}^{J} \left(\det M_j(\xi)\right)^2 $.
Then, from the rank condition based on the submatrix \cite{MatrixAnalysis}, we obtain
\begin{align}
	\label{eq:lemma_1_01}
	\rank U_{(k)}^{(q,T)}\! <\! mq~\!\Leftrightarrow\!~\det M_j(\xi)\!=\!0,~\forall j~\!\Leftrightarrow\!~g(\xi)\!=\!0.
\end{align}
Moreover, since each entry of $ U_{(k)}^{(q,T)} $ depends linearly on $ \xi $, each $ \det M_j(\xi) $ is a multivariate polynomial in $ \xi $, hence $ g(\xi) $ is also a multivariate polynomial.}

{We then claim that $ g $ is not identically zero.
Since $ T \geq mq $, there exists a deterministic input segment $ \xi^* $ such that the corresponding Hankel matrix $ \mathscr{H}_q(\xi^*) $ has full row rank $ mq $.
Indeed, this can be achieved by using an impulse-type input, as shown in \cite[Theorem 2]{2023LCSSAlsalti}.
Therefore, at least one $ mq \times mq $ submatrix of $ \mathscr{H}_q(\xi^*) $ is invertible, and thus $ \det M_j(\xi^*) \neq 0 $ for some $ j $, and thus $ g(\xi^*) > 0 $.
Consequently, $ g $ is not the zero polynomial.}

{
The zero set of a nontrivial polynomial has Lebesgue measure zero \cite[Proposition 1]{Lebesgue_Zero}, namely, the set $ \{\xi: g(\xi) = 0\} $ has measure zero.
Under Assumption~\ref{assumption:input}, $ \xi $ admits a density with respect to the Lebesgue measure, i.e., $ \xi $ is absolutely continuous, because a nondegenerate Gaussian watermark is injected into the nominal control input at each time.
Any Lebesgue-null set of $ \xi $ has probability zero (see, e.g., \cite{Probability}), and thus we have $ \mathbb{P}\left(g(\xi)=0\right)=0 $.
Using (\ref{eq:lemma_1_01}), this implies $ \mathbb{P}(\rank U_{(k)}^{(q,T)}\! <\! mq) = 0 $.
Equivalently, we have $ \mathbb{P}(\rank U_{(k)}^{(q,T)}\! =\! mq) = 1 $, namely, $ \rank U^{(q,T)}_{(k)}  \overset{\mathrm{a.s.}}{=} mq $.
Since the set of indices $ k \in [0,N-T-q+1] $ is finite, taking the intersection over all $ k $ preserves probability one, which concludes the proof.
\hspace*{\fill}~\QED}

\section{Proof of Lemma~\ref{lemma:X_rank}}
\label{appendix:proof_lemma_X}
{
This proof proceeds in a similar manner as in Lemma~\ref{lemma:U_k}.
Fix any $  k \in [0,N-T+1]  $.
For notational simplicity, denote the input data from time $ k $ to $ k+T-2 $ by $ \eta \triangleq u^{[k,k+T-2]}\in\R^{m(T-1)} $.
Under Assumption~\ref{assumption:input}, the random vector $ \eta $ is absolutely continuous and admits a density with respect to the Lebesgue measure.
Now fix an arbitrary realization of the past sequence, i.e., fix the state $ x(k) $ based on the state sequence $ \{x(t)\}_{t=0}^{k-1} $ and the input sequence $ \{u(t)\}_{t=0}^{k-1} $.
From linearity of the system, then, every entry of $ X^{(T)}_{(k)} $ is an affine function of $ \eta $.
Let $ \{R_j(\eta)\}_{j=1}^\mathfrak{J} $ denote the collection of all $ n \times n $ submatrices of $ X^{(T)}_{(k)} $, where $ \mathfrak{J} \triangleq \binom{T}{n}. $
Also, define the scalar function as $ h(\eta) \triangleq \sum_{j=1}^{\mathfrak{J}} \left(\det R_j(\eta)\right)^2 $.
Then, we obtain
\begin{align}
	\label{eq:lemma_2_01}
	\rank X_{(k)}^{(T)}\! <\! n~\Leftrightarrow~\det R_j(\eta)\!=\!0,~\forall j~\Leftrightarrow~h(\eta)\!=\!0.
\end{align}
Since each entry of $ X_{(k)}^{(T)} $ depends linearly on $ \eta $, each $ \det R_j(\eta) $ is a multivariate polynomial in $ \eta $, hence $ h(\eta) $ is also a multivariate polynomial.}

{
Next, we show that $ h $ is not identically zero.
Since the system is controllable, there exists an input segment $ \eta^* $ such that the resulting state sequence $ \{x(t)\}_{t=k+1}^{k+T-1} $ spans $ \R^n $.
Recalling $ T \geq n+1 $, these states appear as columns of $ X^{(T)}_{(k)} $, which implies that at least one $ n \times n $ submatrix of $ X^{(T)}_{(k)} $ is invertible.
Hence, there exists $ \eta^* $ such that $ h(\eta^*) > 0 $, and thus $ h $ is not the zero polynomial.}

{
Since $ h $ is a nonzero polynomial, its zero set has Lebesgue measure zero.
Therefore, we have $ \mathbb{P}\left(h(\eta)=0\right)=0 $, which implies $ \mathbb{P}(\rank X_{(k)}^{(T)}\! =\! n) = 1 $, and thus, $ \rank X^{(T)}_{(k)}  \overset{\mathrm{a.s.}}{=} n $.
Since the set of indices $ k \in [0,N-T] $ is finite, taking the intersection over all $ k $ preserves probability one, which concludes the proof.
\hspace*{\fill}~\QED}

\section{Proof of Theorem~\ref{theorem:clean_attack_identification}}
\label{appendix:proof_of_theorem}

We first outline the proof approach.
Since we do not know which data are clean, for given $ Z^{(q,T)}_{(t)} $ and $ z^{[k, k+q-1]} $, it is impossible to determine whether they belong to the clean, transition, or attack interval.
Therefore, we analyze all possible combinations among the three intervals. Specifically, by examining all combinations {of $ Z^{(q,T)}_{(t)} \in \{\mathrm{clean,transition,attack}\} $ and $ z^{[k, k+q-1]} \in \{\mathrm{clean,transition,attack}\} $, we prove Theorem~\ref{theorem:clean_attack_identification}.
Specifically, under the conditions of Theorem~\ref{theorem:clean_attack_identification}, we derive that, for all $ \Gamma \subseteq \cP$ and for all $ Z^{(q,T)}_{(t)}\in \{\mathrm{clean,transition,attack}\} $, if $ \A^* \subseteq \Gamma $, then
\begin{align}
	\label{eq:proof_theorem_first}
	\gamma^\Gamma_{(t, k)} \overset{\mathrm{a.s.}}{=} 0,~\forall z^{[k, k+q-1]} \!\in\! \{\mathrm{clean,transition,attack}\},
\end{align}
which is sufficient to prove (\ref{eq:theorem_first_eq}).
Additionally, we show that, if $ \A^* \nsubseteq \Gamma $, then there exist $ Z^{(q,T)}_{(t)} $ in the clean interval and $ z^{[k, k+q-1]} $ in the transition interval such that $ \gamma^\Gamma_{(t, k)} \overset{\mathrm{a.s.}}{\neq} 0 $, which is sufficient to prove (\ref{eq:theorem_second_eq}).}
%
%
\subsubsection*{Case 1}
We first consider the case when $ Z^{(q,T)}_{(t)} $ is in the clean interval.
{The following lemma shows that (\ref{eq:proof_theorem_first}) holds when $ Z^{(q,T)}_{(t)} $ is in the clean interval.}
\begin{lemma}
\label{lemma:Z_clean}
{Suppose the same assumptions as in Theorem~\ref{theorem:clean_attack_identification}.}
If $ Z^{(q,T)}_{(t)} $ is in the clean interval, then {(\ref{eq:proof_theorem_first}) holds for all $ \Gamma \subseteq \cP $ such that $ \A^* \subseteq \Gamma $}.
\end{lemma}
\begin{proof}
Since $ Z^{(q,T)}_{(t)} $ is in the clean interval, we have
\begin{align}
	\label{eq:tilde_Z_clean}
	Z^{(q,T )}_{(t)} = \left[
	\begin{array}{c}
		U^{(q,T )}_{(t)} \\ \mathcal{O}^q X^{(T)}_{(t)} + \mathcal{T}^q U^{(q,T )}_{(t)}
	\end{array}\right]
\end{align}
{Now assume that $ T \geq m(q+n)+n $. Then $ T-n \geq m(q+n) $.
	Consider the Hankel matrix of depth $ q+n $ constructed from the same input window, namely,
	\begin{align*}
		U^{(q+n, T-n)}_{(t)} = \mathscr{H}_{q+n}\!\left(u^{[t,t+T+q-2]}\right) \in \R^{m(q+n)\times(T-n)}.
	\end{align*}
	Applying Lemma~\ref{lemma:U_k} with $ (q,T) $ replaced by $ (q+n, T-n) $, $ U^{(q+n, T-n)}_{(t)} $ has full row rank almost surely for all $ t \in[0,N-T-q+1] $.
	Hence, $ u^{[t,t+T+q-2]} $ is persistently exciting of order $ q+n $ almost surely.}
Therefore, from Willems' fundamental lemma (see \cite{2005SCLWillems} or \cite[Ch. 8]{Behavioral}), the Hankel matrix $ Z^{(q,T )}_{(t)} $ constructed from the (clean) I/O data has the correct left kernel of the attack-free system {almost surely}.
In other words, for every I/O vector $ z^{[k, k+q-1]} $ in the clean interval, we have $ K^q_{(t)} z^{[k,k+q-1]} {\overset{\mathrm{a.s.}}{=}} 0 $, where $ K^q_{(t)} $ is the kernel representation of the attack-free system obtained through the SVD of (\ref{eq:SVD}).
From (\ref{eq:gamma}), this implies $ \gamma^\Gamma_{(t, k)}  = P^\Gamma_{(t)} K^q_{(t)} z^{[k,k+q-1]} {\overset{\mathrm{a.s.}}{=}}  0 $
for all $ \Gamma \subseteq \cP $ and for all $ z^{[k, k+q-1]} $ in the clean interval.

Then, consider $ z^{[k, k+q-1]} $ in the transition or attack interval.
For each $ z^{[k, k+q-1]} $, decompose
\begin{align}
	\label{eq:lemma_zeta}
	z^{[k,k+q-1]} = \zeta^{[k, k+q-1]} + \left[
	\begin{array}{c}
		0 \\ a^{[k,k+q-1]}
	\end{array}\right],
\end{align}
where $ \zeta^{[k, k+q-1]} $ denotes an unknown \textit{attack-free} stacked I/O vector, which is defined as
\begin{align*}
	\zeta^{[k,k+q-1]} \triangleq \left[
	\begin{array}{c}
		u^{[k,k+q-1]} \\ \mathcal{O}^q x(k) +\mathcal{T}^q u^{[k,k+q-1]}
	\end{array}\right]\in \R^{mq+pq}.
\end{align*}
Since now $ K^q_{(t)} $ is the kernel representation of the attack-free system, we obtain $ K^q_{(t)}\zeta^{[k,k+q-1]} {\overset{\rm a.s.}{=}} 0 $, and thus,
\begin{align}
	\label{eq:lemma_gamma}
	\gamma^\Gamma_{(t, k)}  = P^\Gamma_{(t)} K^q_{(t)} z^{[k,k+q-1]}{\overset{\rm a.s.}{=} P^\Gamma_{(t)} K^q_{(t)}}\left[\!\!
	\begin{array}{c}
		0 \\ a^{[k,k+q-1]}
	\end{array}\!\!\right].
\end{align}
From the construction of $ P^\Gamma_{(t)} $, it follows that
\begin{align*}
	P_{(t)}^\Gamma K^{q}_{(t)} \!\left[\!\!
	\begin{array}{c}
		0 \\ v
	\end{array}\!\!\right]  =P_{(t)}^\Gamma Q^2_{(t)} v = 0,~\forall v \in \im \mathbb{I}^\Gamma_q.
\end{align*}
Since $ \A^* \subseteq \Gamma $, the stacked attack vector follows $ a^{[k,k+q-1]} \in \im \mathbb{I}^{\A^*}_q \subseteq \im \mathbb{I}^{\Gamma}_q $, which implies
\begin{align}
	\label{eq:lemma_a_k}
	P_{(t)}^\Gamma K^{q}_{(t)} \!\left[\!\!
	\begin{array}{c}
		0 \\ a^{[k,k+q-1]}
	\end{array}\!\!\right] = P_{(t)}^\Gamma Q^2_{(t)} a^{[k,k+q-1]}   = 0.
\end{align}
{
	Therefore, from (\ref{eq:lemma_gamma}), we obtain $ \gamma^\Gamma_{(t, k)} \overset{\mathrm{a.s.}}{=}  0 $ for all $ \Gamma $ such that $ \mathcal{A}^* \subseteq \Gamma $ and for all $ z^{[k, k+q-1]} $ in the transition or attack interval.
	Consequently, (\ref{eq:proof_theorem_first}) holds for all $ \Gamma \subseteq \cP $ such that $ \A^* \subseteq \Gamma $}.
\end{proof}

{We also introduce the following lemma to prove that (\ref{eq:theorem_second_eq}) holds when $ Z^{(q,T)}_{(t)} $ is in the clean interval and $ z^{[k, k+q-1]} $ is in the transition interval.
\begin{lemma}
	\label{lemma:Z_clean_noteq}
	Suppose the same premise as in Theorem~\ref{theorem:clean_attack_identification}.
	If $ \A^* \nsubseteq \Gamma $, then there exist $ Z^{(q,T)}_{(t)} $ in the clean interval and $ z^{[k, k+q-1]} $ in the transition interval such that $ \gamma^\Gamma_{(t, k)} \overset{\mathrm{a.s.}}{\neq} 0 $.
\end{lemma}}
\begin{proof}
From Lemma~\ref{lemma:Z_clean}, if $ T \geq m(q+n)+n $, $ K^q_{(t)} $ is the kernel representation of the attack-free system.
Utilizing the result of \cite{2014DingAutomatica}, $ K^q_{(t)} $ satisfies
\begin{align*}
	K^q_{(t)}
	= \left[
	\begin{array}{cc}
		Q^1_{(t)} & Q^2_{(t)}
	\end{array}\right]
	= \left[
	\begin{array}{cc}
		-(\mathcal{O}^q)^\bot \mathcal{T}^q & (\mathcal{O}^q)^\bot
	\end{array}\right],
\end{align*}
where $ (\mathcal{O}^q)^\bot $ is the orthogonal complement matrix of $ \mathcal{O}^q $ (i.e., $ (\mathcal{O}^q)^\bot \mathcal{O}^q = 0 $).
Then, from (\ref{eq:lemma_gamma}), it follows that $ \gamma^\Gamma_{(t,k)} \overset{\rm a.s.}{=} P^\Gamma_{(t)} (\cO^q)^\bot a^{[k,k+q-1]} $.
{
	In this proof, we show that $ P_{(t)}^\Gamma (\mathcal{O}^q)^\bot a^{[k,k+q-1]} \overset{\rm a.s.}{\neq} 0,\exists k,t $ by contradiction.
	To this end, assume that $ P_{(t)}^\Gamma (\mathcal{O}^q)^\bot a^{[k,k+q-1]} \overset{\rm a.s.}{=} 0,\forall k,t $.
	Similar to the proof of Proposition~\ref{proposition:detection_attack}, without loss of generality, consider $ k^\sharp \triangleq k_0 + \tau- q + 1 $, which is the first time of the transition interval after $ \K_0 $ in the vector sense (cf.~Fig.~\ref{fig:Relationship_Lambda}).
	Define
	\begin{align}
		\label{eq:a_sharp}
		\psi \triangleq a^{[k^\sharp, k^\sharp + q -1]} = \left[\!~0^\top~\cdots~0^\top~a(k_0 + \tau)^\top~\!\right]^\top \!\!\in \R^{pq}.
	\end{align}
	Given that the construction of $ P_{(t)}^\Gamma $, $ P_{(t)}^\Gamma (\mathcal{O}^q)^\bot \psi \overset{\rm a.s.}{=} 0 $ implies
	\begin{align}
		\label{eq:lemma_im}
		\left(\cO^q\right)^\bot \psi \in \im \left(\left(\cO^q\right)^\bot \mathbb{I}_q^\Gamma \right),~\text{almost surely}.
	\end{align}
	This implies that there exists a vector $ b \in \R^{|\Gamma|q} $ such that
	\begin{align*}
		\left(\cO^q\right)^\bot \psi \overset{\rm a.s.}{=} \left(\cO^q\right)^\bot \mathbb{I}_q^\Gamma b
		~\Leftrightarrow~\left(\cO^q\right)^\bot \left(\psi - \mathbb{I}_q^\Gamma b\right) \overset{\rm a.s.}{=} 0,
	\end{align*}
	namely, there exist $ b \in \R^{|\Gamma|q} $ and $ x\in \R^n $ such that $ \psi - \mathbb{I}_q^\Gamma b \overset{\rm a.s.}{=} \cO^q x $.
	Recalling the structures of $\psi$, $ \mathbb{I}_q^\Gamma $, and $ \cO^q $, we have
	\begin{align}
		\label{eq:a_sharp_lemma_01}
		\left[
		\begin{array}{c}
			0 \\ \vdots \\ 0 \\ a(k_0 + \tau)
		\end{array}\right] - \left[
		\begin{array}{c}
			I_p^\Gamma b_1 \\ \vdots \\ I_p^\Gamma b_{q-1} \\  I_p^\Gamma b_{q}
		\end{array}\right] \overset{\rm a.s.}{=} \left[
		\begin{array}{c}
			C \\ \vdots \\CA^{q-2} \\ CA^{q-1}
		\end{array}\right]x
	\end{align}
	where $ b_i \in \R^{|\Gamma|} $ denotes the $ i $th block of $ b $.
	Since now $  \A^* \nsubseteq \Gamma $, there exists a nonempty set $  \A^\complement \triangleq \A^* \setminus \Gamma $.
	Then, since the submatrix of $ I_p^\Gamma $ whose rows indexed by $ \A^\complement $ is zero, from (\ref{eq:a_sharp_lemma_01}), we obtain $ C_{\A^\complement} x ,\ldots, C_{\A^\complement} A^{q-2}x \overset{\rm a.s.}{=} 0 $, where $ C_{\A^\complement} \in \R^{|\A^\complement|\times n} $ denotes the submatrix obtained from $ C $ by removing all rows except those indexed by $ \A^\complement $.
	Recalling $ q \geq n+1 $, from the Cayley-Hamilton theorem, it follows that $ C_{\A^\complement}A^{q-1} x \overset{\rm a.s.}{=} 0 $.
	Therefore, from (\ref{eq:a_sharp_lemma_01}), it must hold that $   a_{\A^\complement}(k_0 + \tau) \overset{\rm a.s.}{=} 0  $, where $   a_{\A^\complement}(k_0 + \tau) \in \R^{|\A^\complement|} $ is the subvector of $ a(k_0 + \tau) $ indexed by $ \A^\complement $.
	However, this contradicts the assumption that $ \supp{a(k_0 + \tau)}= \A^* $.
	Consequently, (\ref{eq:lemma_im}) does not hold, which implies $ P_{(t)}^\Gamma (\mathcal{O}^q)^\bot \psi \!\overset{\rm a.s.}{\neq} \!0 $.
	This concludes the proof.
}
\end{proof}

\subsubsection*{Case 2}
We next address the case when $ Z^{(q,T)}_{(t)} $ is in the transition interval.
\begin{lemma}
\label{lemma:Z_attack_transition}
Suppose the same premise as in Theorem~\ref{theorem:clean_attack_identification}.
If $ Z^{(q,T)}_{(t)} $ is in the transition interval, then {(\ref{eq:proof_theorem_first}) holds for all $ \Gamma \subseteq \cP $ such that $ \A^* \subseteq \Gamma $}.
\end{lemma}
\begin{proof}
As with Lemma~\ref{lemma:Z_clean}, for each $ k $, we decompose $ z^{[k,k+q-1]} $ as (\ref{eq:lemma_zeta}) by using the unknown attack-free stacked I/O vector $ \zeta^{[k, k+q-1]} $.
%
Since $ \A^* \subseteq \Gamma $, (\ref{eq:lemma_a_k}) holds, and thus we obtain
\begin{align}
	\gamma^\Gamma_{(t, k)}  =P^\Gamma_{(t)} K^q_{(t)} z^{[k,k+q-1]} = P^\Gamma_{(t)} K^q_{(t)} \zeta^{[k,k+q-1]}.
\end{align}

{We next show $ P^\Gamma_{(t)} K^q_{(t)} \zeta^{[k,k+q-1]} \overset{\rm a.s.}{=} 0 $.}
From the SVD of (\ref{eq:SVD}), we obtain
\begin{align*}
	&  K^{q}_{(t)}\left[
	\begin{array}{c}
		U^{(q,T)}_{(t)} \\ \mathcal{O}^q X^{(T)}_{(t)} +\mathcal{T}^q U^{(q,T)}_{(t)}+ \Lambda^{(q,T)}_{(t)}
	\end{array}\right] =  0 \\
	\Rightarrow & K^{q}_{(t)}\underbrace{\left[
		\begin{array}{c}
			U^{(q,T)}_{(t)} \\ \mathcal{O}^q X^{(T)}_{(t)}  + \mathcal{T}^q U^{(q,T)}_{(t)}
		\end{array}\right]}_{{\widetilde Z^{(q,T)}_{(t)}}} =  -K^{q}_{(t)}\left[
	\begin{array}{c}
		0\\ \Lambda^{(q,T)}_{(t)}
	\end{array}\right],
\end{align*}
{where $ \widetilde Z^{(q,T)}_{(t)} $ denotes the \textit{attack-free} I/O Hankel matrix.
	Multiplying $ P^\Gamma_{(t)} $, we have
	\begin{align*}
		P^\Gamma_{(t)} K^q_{(t)}\widetilde Z^{(q,T)}_{(t)} = - P^\Gamma_{(t)}K^q_{(t)}\left[\!\!
		\begin{array}{c}
			0\\ \Lambda^{(q,T)}_{(t)}
		\end{array}\!\!\right]= -P^\Gamma_{(t)} Q^2_{(t)} \Lambda^{(q,T)}_{(t)}.
	\end{align*}
	Recalling (\ref{eq:lemma_a_k}), it also follows that $ P_{(t)}^\Gamma Q^2_{(t)} a^{[t,t+q-1]}   = 0 $ for all $ t $.
	Thus, we have $  -P^\Gamma_{(t)} Q^2_{(t)} \Lambda^{(q,T)}_{(t)} = 0 $, which implies $ P^\Gamma_{(t)} K^q_{(t)}\widetilde Z^{(q,T)}_{(t)} = 0 $.
	This yields $ \mathrm{row}(P^\Gamma_{(t)} K^q_{(t)}) \subseteq \ker (\widetilde Z^{(q,T)}_{(t)})^\top $, where $ \mathrm{row}(\cdot) $ denotes the row space of a matrix.}

{Now $ \widetilde Z^{(q,T)}_{(t)} $ is the attack-free Hankel matrix.
	From the proof of Lemma~\ref{lemma:Z_clean}, under the same conditions, Willems' fundamental lemma implies that $ \widetilde Z^{(q,T)}_{(t)} $ has the correct left kernel of the attack-free system almost surely.
	Therefore, using the result of \cite{2014DingAutomatica} again, we obtain $ \ker (\widetilde Z^{(q,T)}_{(t)})^\top \overset{\rm a.s.}{=} \mathrm{row}(\left[-(\mathcal{O}^q)^\bot \mathcal{T}^q ~~(\mathcal{O}^q)^\bot\right]) $,
	and thus
	\begin{align}
		\mathrm{row}\left(P^\Gamma_{(t)} K^q_{(t)}\right) \overset{\rm a.s.}{\subseteq}\mathrm{row}\left(\left[
		\begin{array}{cc}
			-(\mathcal{O}^q)^\bot \mathcal{T}^q & (\mathcal{O}^q)^\bot
		\end{array}\right]\right),
	\end{align}
	which implies that there exists a time-varying matrix $ R_{(t)} $ such that
	\begin{align}
		P^\Gamma_{(t)} K^q_{(t)} \overset{\rm a.s.}{=} R_{(t)}\left[
		\begin{array}{cc}
			-(\mathcal{O}^q)^\bot \mathcal{T}^q & (\mathcal{O}^q)^\bot
		\end{array}\right].
	\end{align}
	Hence, we obtain
	\begin{align}
		&P^\Gamma_{(t)} K^q_{(t)} \zeta^{[k,k+q-1]} \nonumber\\
		& \overset{\rm a.s.}{=} R_{(t)}\left[\!\!
		\begin{array}{cc}
			-(\mathcal{O}^q)^\bot \mathcal{T}^q & (\mathcal{O}^q)^\bot
		\end{array}\!\!\right]\zeta^{[k,k+q-1]} = 0,
	\end{align}
	which implies (\ref{eq:proof_theorem_first}) holds for all $ \Gamma \subseteq \cP $ such that $ \A^* \subseteq \Gamma $.}
\end{proof}

\subsubsection*{Case 3}
We finally deal with the case when $ Z^{(q,T)}_{(t)} $ is in the attack interval.
\begin{lemma}
\label{lemma:Z_attack}
Suppose the same premise as in Theorem~\ref{theorem:clean_attack_identification}.
If $ Z^{(q,T)}_{(t)} $ is in the attack interval, then {(\ref{eq:proof_theorem_first}) holds for all $ \Gamma \subseteq \cP $ such that $ \A^* \subseteq \Gamma $}.
\end{lemma}
\begin{proof}
The same proof of Lemma~\ref{lemma:Z_attack_transition} can be applied, and thus, {(\ref{eq:proof_theorem_first}) holds for all $ \Gamma \subseteq \cP $ such that $ \A^* \subseteq \Gamma $.}
\end{proof}

Combining the results of these lemmas, Theorem~\ref{theorem:clean_attack_identification} follows directly.

{\noindent\hspace{2em}{\itshape Proof of Theorem~\ref{theorem:clean_attack_identification}:}}
From Lemmas \ref{lemma:Z_clean}, \ref{lemma:Z_attack_transition}, and \ref{lemma:Z_attack}, regardless of whether $ Z^{(q,T)}_{(t)} $ is in the clean, transition, or attack interval, {(\ref{eq:proof_theorem_first}) holds for all $ \Gamma \subseteq \cP $ such that $ \A^* \subseteq \Gamma $, which implies (\ref{eq:theorem_first_eq}) holds.
Additionally, from Lemma~\ref{lemma:Z_clean_noteq}, if $ \A^* \nsubseteq \Gamma$, then there exist $ k \in [0,N-q] $ and $ t \in [0,N-T-q+1] $ such that $ \gamma^\Gamma_{(t, k)} \! \overset{\mathrm{a.s.}}{\neq} \!0 $, which implies (\ref{eq:theorem_second_eq}) holds.}
\hspace*{\fill}~\QED

\vspace{-8mm}

\begin{IEEEbiography}{Takumi Shinohara} (Member, IEEE) received the B.E., M.E., and Ph.D. degrees from Keio University, Tokyo, Japan, in 2016, 2018, and 2024, respectively. From 2018 to 2025, he was a consultant at Mitsubishi Research Institute, and from 2024 to 2025, he was a visiting researcher at Keio University. Since 2025, he has been a Postdoctoral Researcher with the Division of Decision and Control Systems at KTH Royal Institute of Technology, Stockholm, Sweden.
His research interests include control system security and secure state estimation.
\end{IEEEbiography}

\vspace{-8mm}

\begin{IEEEbiography}{Karl Henrik Johansson} (Fellow, IEEE) received the M.Sc. degree in electrical engineering and the Ph.D. degree in automatic control from Lund University, Lund, Sweden, in 1992 and 1997, respectively.

He is a Swedish Research Council Distinguished Professor in electrical engineering and computer science with the KTH Royal Institute of Technology, Stockholm, Sweden, and the Founding Director of Digital Futures. He has held Visiting Positions with UC Berkeley, Caltech, NTU, and other prestigious institutions. His research interests include networked control systems and cyber-physical systems with applications in transportation, energy, and automation networks.

Dr. Johansson was the recipient of numerous best paper awards and various distinctions from IEEE, IFAC, and other organizations, for his scientific contributions, and also Distinguished Professor by the Swedish Research Council, Wallenberg Scholar by the Knut and Alice Wallenberg Foundation, Future Research Leader by the Swedish Foundation for Strategic Research, triennial IFAC Young Author Prize and IEEE CSS Distinguished Lecturer, and 2024 IEEE CSS Hendrik W. Bode Lecture Prize. His extensive service to the academic community includes being President of the European Control Association, IEEE CSS Vice President Diversity, Outreach \& Development, and Member of IEEE CSS Board of Governors and IFAC Council. He was on the editorial boards of Automatica, IEEE Transactions on Automatic Control, IEEE Transactions on Control of Network Systems, and many other journals. He has also been a Member of the Swedish Scientific Council for Natural Sciences and Engineering Sciences. He is Fellow of the Royal Swedish Academy of Engineering Sciences.
\end{IEEEbiography}

\vspace{-8mm}

\begin{IEEEbiography}{Henrik Sandberg} (Fellow, IEEE) is Professor at the Division of Decision and Control Systems, KTH Royal Institute of Technology, Stockholm, Sweden. He received the M.Sc. degree in engineering physics and the Ph.D. degree in automatic control from Lund University, Lund, Sweden, in 1999 and 2004, respectively. From 2005 to 2007, he was a Postdoctoral Scholar at  the California Institute of Technology, Pasadena, USA. In 2013, he was a Visiting Scholar at the Laboratory for Information and Decision Systems (LIDS) at MIT, Cambridge, USA. He has also held visiting appointments at the Australian National University and the University of Melbourne, Australia. His current research interests include security of cyber-physical systems, power systems, model reduction, and fundamental limitations in control. Dr. Sandberg was a recipient of the Best Student Paper Award from the IEEE Conference on Decision and Control in 2004, an Ingvar Carlsson Award from the Swedish Foundation for Strategic Research in 2007, and a Consolidator Grant from the Swedish Research Council in 2016. He has served on the editorial boards of IEEE Transactions on Automatic Control and the IFAC Journal Automatica. He is Fellow of the IEEE.
\end{IEEEbiography}

\end{document}